\documentclass[journal]{IEEEtran}
\usepackage[cmex10]{amsmath}
\usepackage{amssymb,amsfonts}
\usepackage{graphicx}
\usepackage{booktabs}

\usepackage{xcolor}
\newif\ifshowrevisions
\showrevisionsfalse
\definecolor{revisionolive}{RGB}{110,130,35}

\newcommand{\rev}[1]{%
  \ifshowrevisions
    \textcolor{revisionolive}{#1}%
  \else
    #1%
  \fi
}

\usepackage{tikz}
\usepackage{algorithm}
\usepackage{algpseudocode}
\usepackage[caption=false,font=footnotesize]{subfig} % replace subcaption
\usepackage{cite}
\usepackage{url}
\usepackage[hidelinks]{hyperref} % keep last

\begin{document}

% TITLE
\title{xHAP: Cross-Modal Attention for Haptic Feedback Estimation in the Tactile Internet}

% AUTHOR INFORMATION
\author{Georgios~Kokkinis,~\IEEEmembership{Graduate~Student~Member,~IEEE}, 
        Alexandros~Iosifidis,~\IEEEmembership{Senior~Member,~IEEE}, 
        and~Qi~Zhang,~\IEEEmembership{Senior~Member,~IEEE}%
\thanks{G. Kokkinis and Q. Zhang are with DIGIT and Department of Electrical and Computer Engineering, Aarhus University, Aarhus, Denmark (e-mail: \{gkokkinis, qz\}@ece.au.dk).}%
\thanks{A. Iosifidis is with the Data Science Research Centre, Tampere University, Tampere, Finland (e-mail: alexandros.iosifidis@tuni.fi).}%
\thanks{This research was supported by the TOAST project, funded by the European Union’s Horizon Europe research and innovation program under the Marie Skłodowska-Curie Actions Doctoral Network (Grant Agreement No. 101073465), the Danish Council for Independent Research project eTouch (Grant No. 1127- 00339B) and NordForsk Nordic University Cooperation on Edge Intelligence (Grant No. 168043).}%
}

% The paper headers
\markboth{IEEE Transactions on Mobile Computing,~Vol.~XX, No.~X, Month~2025}%
{Kokkinis \MakeLowercase{\textit{et al.}}: Cross-Modal Attention for Haptic Feedback Estimation}

% Make the title area
\maketitle

% ABSTRACT & KEYWORDS
\begin{abstract}
The Tactile Internet requires ultra-low latency and high-fidelity haptic feedback to enable immersive teleoperation. A key challenge is to ensure ultra-reliable and low-latency transmission of haptic packets under channel variations and potential network outages. To address these issues, one approach relies on local estimation of haptic feedback at the operator side. However, designing an accurate estimator that can faithfully reproduce the true haptic forces remains a significant challenge. In this paper, we propose a novel deep learning architecture, xHAP, based on cross-modal attention to estimate haptic feedback. xHAP fuses information from two distinct data streams: the teleoperator's historical force feedback and the operator's control action sequence. We employ modality-specific encoders to learn temporal representations, followed by a cross-attention layer where the teleoperator haptic data attend to the operator input. This fusion allows the model to selectively focus on the most relevant operator sensory data when predicting the teleoperator's haptic feedback. The proposed architecture reduces the mean-squared error by more than two orders of magnitude compared to existing methods and lowers the SNR requirement for reliable transmission by $10~\mathrm{dB}$ at an error threshold of $0.1$ in a 3GPP UMa scenario. Additionally, it increases coverage by $138\%$ and supports $59.6\%$ more haptic users even under 10 dB lower SNR compared to the baseline.
\end{abstract}


\IEEEpeerreviewmaketitle

\section{Introduction}
\IEEEPARstart{T}{he} Tactile Internet (TI) aims to revolutionize human-machine interaction by enabling real-time remote teleoperation with haptic feedback. In typical haptic teleoperation, a human operator transmits control commands to a remote robotic system (teleoperator). Haptic feedback consists of kinesthetic (force, motion) and tactile (vibration, texture) information, which is transmitted back to the operator, creates a closed-loop system to provide task precision and a sense of telepresence. However, a critical bottleneck of the control loop is the unreliable transmission of packets between local and remote environments. Communication outages lead to desynchronization and system instability, thereby compromising transparency, immersiveness, and operational safety. Consequently, developing robust control methodologies is imperative to maintain performance under unreliable transmission conditions.

\begin{figure*}[!t]
 \centering
 \includegraphics[width=1.0\textwidth]{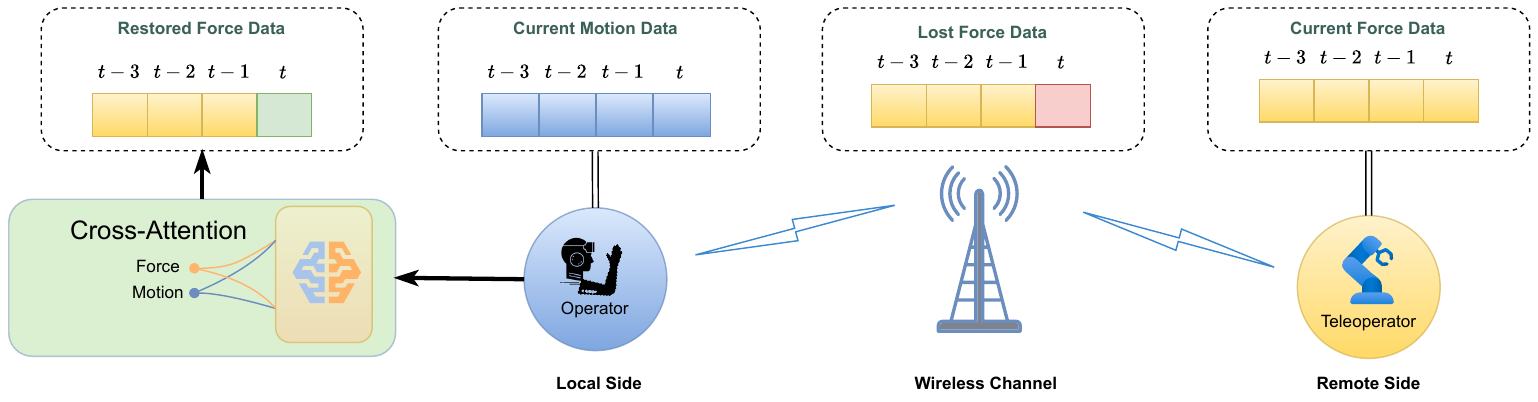}
 \caption{Diagram of the packet estimation pipeline. \rev{The yellow packets indicate known force values from teleoperator; the red packet indicates the lost force value; the blue packets indicate known command-signal packets from operator; and the green packet indicates the restored force value.}}
 \label{fig:overall_diagram}
\end{figure*}

From a technical perspective, the primary challenge in realizing such systems is the extreme quality-of-service (QoS) demands of the communication channel. For the system to feel transparent and remain stable, the round-trip time (RTT) for haptic signals must be approximately 1 millisecond~\cite{fettweis2014}. Although from a neurophysiological perspective humans can compensate for higher latency, the haptic loop \rev{can become unstable} under significant delay. This ``1 ms Challenge" represents a significant leap from the latency of current networks. Furthermore, mission-critical applications like remote surgery demand ultra-high reliability, often exceeding 99.999\%, to prevent catastrophic failures from packet loss~\cite{antonakoglou2018,popovski2018urllc}. In recent years, the quantification of TI application performance has emerged as a significant research priority~\cite{kroep2024}, and Quality of Experience (QoE) remains highly variable across individual users.

To address the highly dynamic challenges inherent in future TI communications, data-driven approaches have utilized predictive models to proactively satisfy stringent haptic requirements. Specifically, deep learning (DL) architectures have become a dominant solution for the estimation and forecasting of haptic time-series data. Over the past decade, state-of-the-art (SotA) DL models perform favorably in estimating non-linear trajectories under strictly time-critical conditions~\cite{zhou2021informer,nie2023patchtst}, compared to conventional methods. In multimodal haptic systems, deep neural networks (DNNs) can be trained on large datasets to extract meaningful features and capture cross-modal correlations. This facilitates the modeling of user intent, which is particularly advantageous for mitigating the impact of network-induced delays \cite{kroep2025}. \rev{The inductive bias of cross-attention in xHAP, that is, its ability to filter and attend only to the relevant operator inputs, provides a lightweight implementation of a Deep Neural Network (DNN), thus enabling real-time force feedback estimation.} To this end, we propose \rev{xHAP}, a dual-branch cross-attention model that enables selective information exchange between teleoperator forces and operator control signals. In essence, cross-attention is a filter that dynamically determines which aspects of the operator's actions are most informative for estimating the teleoperator's next state~\cite{li2023blip2}. Throughout the paper, all force values and force-related error metrics are expressed in Newtons (N). 

% --- NEW PARAGRAPH ADDED HERE TO ADDRESS REVIEWER 3, COMMENT 1 ---
\rev{This paper focuses on kinesthetic force feedback, rather than tactile cues such as vibration or texture. In bilateral force-feedback systems, kinesthetic signals are commonly sampled and transmitted at approximately $1~\mathrm{kHz}$, corresponding to about $1000$ haptic packets per second, to maintain stable and transparent interaction~\cite{aijaz2016realizing}. At the communication level, Tactile Internet and URLLC studies typically associate such scenarios with end-to-end latency on the order of $1~\mathrm{ms}$ and reliability around $99.999\%$~\cite{fettweis2014,sharma2020towards}. By contrast, tactile cues are often treated as cutaneous signals such as vibrotactile feedback. For example, the IEEE 1918.1 haptic-codec framework discusses single-point tactile coding for a $1$D vibrotactile signal, e.g., $100~\mathrm{Hz}$ with $16$-bit samples~\cite{IEEEP1918}. Thus, this work targets kinesthetic force feedback because its packet loss and latency constraints are directly tied to closed-loop stability and transparency.}

% ---------------------------------------------------------------

In closed-loop teleoperation, the control signals of the operator are strongly correlated with the force feedback of the teleoperator. It is reasonable to assume that cross-modal attention is a fitting option for selectively attending from one input sequence to the other. In this paper, our aim is to utilize cross-attention between teleoperator and operator time-series to estimate force feedback, with the purpose of recovering lost packets during network outages. The overall framework is illustrated in Fig.~\ref{fig:overall_diagram}. \rev{By selectively attending to the operator signals most relevant to the current force estimate, xHAP achieves a lightweight DNN implementation suitable for real-time force feedback estimation.} In particular, the main contributions of this paper are the following:
\begin{itemize}
    \item \textit{Haptic Cross-attention:} We propose xHAP, a cross-modal attention architecture for the DL estimator, which selectively attends to the operator's input. This selective filter method is fitting for lightweight models for real-time inference. Specifically, we use two separate branches for the operator and teleoperator modalities, where each branch starts with the input sequence, and is encoded through a GRU layer. From the output of the two GRUs, a cross-attention layer fuses the learnt data representations. Finally, the fused representation is transformed by a linear layer. This model performs better than more complex models and other lightweight estimators, yielding an average mean squared error (MSE) of $2.12\times10^{-4}$.

    \item \textit{Force estimation and autoregressive restoration:}
    Our experiments use real-world haptic traces and multi-modal data to train deep neural networks (DNNs) for force estimation under a wireless channel model with potential packet loss. To address missing data, we propose an autoregressive restoration approach that leverages previous force feedback along with current position and velocity signals. When consecutive packets are lost, each newly estimated force value is recursively used for the next prediction, enabling accurate, continuous force estimation and allowing direct comparison between scenarios with and without restoration.
    
    \item \textit{xHAP for enhanced reliability:}  
    With strict estimation error requirement across all tasks (threshold of $0.1$), the proposed restoration method substantially enhances the reliability of wireless haptic communication by reconstructing lost or delayed packets at the operator side. This reduces the effective packet loss rate in the joint communication–control loop. Compared to the no-restoration baseline, xHAP lowers the required SNR by $10.58~\mathrm{dB}$ under 3GPP UMa settings, extends coverage by $138\%$, and supports $59.6\%$ more haptic users, achieved at an SNR that is $10~\mathrm{dB}$ lower than the no-restoration baseline.

\end{itemize}

The remainder of this paper is organized as follows: in Section~\ref{sec:wireless}, we describe a wireless channel and packet error model used to simulate network outage conditions. We describe the xHAP cross-attention estimator in section~\ref{sec:estimator}, providing details about the architectural biases implemented in the structure. Building on the structural analysis, Section~\ref{sec:comparison} evaluates our estimator's performance relative to other DL models. We also analyze how the features of haptic traces correlate with the performance of each estimator. In section~\ref{sec:results}, we showcase multiple experiments that quantitatively demonstrate the value of our estimator, comparing the reliability of the restoration scenario with DL against baseline no-restoration. Finally in section~\ref{sec:conclusion} we give our concluding statements.

\section{Related Work}

In recent years, various methods have been investigated to enhance the reliability of haptic data transmission. Within the vision of the TI, services involving haptic communication can multiplex URLLC packets with enhanced Mobile Broadband (eMBB) resources~\cite{Kim2019}, enabling more reliable and efficient resource scheduling. Such methods can be realized through the multiplexing of different numerologies in 5G New Radio. For instance, in~\cite{Lim2018}, users with similar mobility characteristics are grouped and allocated to the same OFDM subband, where each group is assigned a specific numerology according to its service requirements. In the context of TI, the video–tactile multiplexing schemes proposed in~\cite{Gokhale2023} have been shown to reduce latency in Wi-Fi–enabled TI systems. Furthermore,~\cite{Durisi2016} presents an information-theoretic approach for optimizing reliability in short-packet transmission, providing valuable insights for URLLC in 5G and beyond. To achieve low latency and high reliability for such packets, techniques such as mini-slotting and packet puncturing are employed to adjust the transmission time interval (TTI) and manage resource allocation according to the targeted service~\cite{Popovski2018}.

In haptic systems, predictive models are frequently used to estimate physical properties, anticipate sensory feedback, and compensate for network outages. In \cite{Navarro2022}, a framework for operating remote surgery is proposed, with its foundation built upon predictive haptic methods. In the field of autonomous driving~\cite{Hosseini2016}, a two-stage predictive framework is proposed to compensate for communication delays through smooth haptic feedback, ensuring the human remains the primary vehicle controller.  Building on the importance of haptics for autonomous systems, \cite{Gao2016} explores tactile understanding for robots by classifying surfaces using visual and physical data. This model, inspired by human cognition, uses DNNs to predict haptic properties and shows that unifying visual and physical signals leads to superior performance over methods with hand-designed features.

In human-in-the-loop teleoperation,~\cite{Xiyan2016} introduce an adaptive estimator with coefficient updates, yielding smooth 1 kHz haptic feedback via sampling and interpolation, though deterministic methods deteriorate over long horizons due to haptic nonlinearities. A data-driven alternative is explored in~\cite{Xu2020}, where RemedyLSTM outperforms linear estimators in packet prediction and resilience to transmission errors. Furthermore, ~\cite{Kizilkaya2023} demonstrate that deep learning models trained with GAN-augmented data achieve higher accuracy while relaxing delay bounds, facilitating flexible resource allocation for ultra-reliable, low-latency teleoperation. In~\cite{Chen2024}, a cross-modal predictive framework is used to reconstruct the visual,audio, and haptic signals over a fitting future horizon.

Even when specifically trained on teleoperation data, the architecture of the models in prior work remains generic. Given the multi-modal nature of the haptic loop, strong correlations can be captured between the command signal channels and force feedback with the use of cross-attention. Cross-attention mechanisms have been widely adopted in recent works for modeling interactions across different modalities and tasks. For instance, the Multi-Modality Cross Attention Network has been proposed to enhance image--sentence matching by effectively capturing semantic alignments between visual and textual representations~\cite{Wei2020}. Similarly, cross-modal self-attention networks have demonstrated strong performance in referring image segmentation by enabling fine-grained reasoning between visual regions and language expressions~\cite{chumachenko2022}. Beyond vision--language applications, cross-attention has also proven effective in natural language processing, where it has been leveraged to adapt pretrained transformers for machine translation, further showcasing its versatility and generalization ability across domains~\cite{gheini2021}. In~\cite{chenVTT2023} Visuo-Tactile Transformers use cross/self-attention to fuse tactile with vision, improving representation learning for manipulation and planning. However, to our knowledge, cross-attention has not been utilized between the input streams of teleoperator and operator. Hence, we propose this method to estimate and restore haptic packets that are lost during wireless transmission.

% Helpers for tidy 4-column notation tables
\newcommand{\pair}[4]{\(#1\) & #2 & \(#3\) & #4\\}
\newcommand{\single}[2]{\multicolumn{4}{@{}l@{}}{\(\mathbf{#1}\)\quad #2}\\}

\begin{table}[!t]
  \caption{Notation}
  \label{tab:notation}
  \centering
  \renewcommand{\arraystretch}{1.15}
  \resizebox{\columnwidth}{!}{%
  \begin{tabular}{p{1.7cm}p{4.1cm}|p{1.7cm}p{4.1cm}}
    \hline
    \textbf{Symbol} & \textbf{Meaning} & \textbf{Symbol} & \textbf{Meaning} \\
    \hline
    $\mu$ & average SNR (dB) &
    $\sigma_{\mathrm{sh}}$ & std.\ dev.\ of shadowing (dB) \\

    $\rho$ & temporal correlation coefficient &
    $Z_t$ & composite shadowing{+}fading (dB) \\

    $S_t,F_t$ & shadowing / fading components &
    $\mathrm{SNR}_t$ & instantaneous SNR at time $t$ (dB) \\

    $G_{\mathrm{FEC}}$ & FEC coding gain (dB) &
    $\mathrm{SNR}^{\mathrm{eff}}_t$ & effective SNR after FEC (dB) \\

    $\gamma_t$ & linear effective SNR, $10^{\mathrm{SNR}^{\mathrm{eff}}_t/10}$ &
    $\mathrm{BER}(\gamma)$ & instantaneous bit error rate \\

    $\mathrm{PER}_t$ & packet error rate at time $t$ &
    $N_b$ & packet size (bits) \\

    $\mathrm{PLR}$ & packet loss rate over horizon &
    $N_{\mathrm{LP}}$ & \# lost packets in horizon \\

    $b(\mathsf{M})$ & bits/symb.\ for modulation $\mathsf{M}$ &
    $\eta$ & spectral efficiency (bits/s/Hz) \\

    $f_s$ & symbol rate &
    $\mathcal{B}$ & bandwidth (Hz) \\

    $R$ & code rate &
    $R_{\mathrm{c}}$ & coded data rate $\,{=}\,\eta\,\mathcal{B}$ \\

    $R_{\mathrm{eff}}$ & goodput / effective throughput &
    $\gamma_{\mathrm{eff}}$ & post-combining SNR (if diversity) \\

    $L_{\mathrm{div}}$ & diversity order (branches) &
    & \\
    \hline

    $d$ & BS--UE distance (m) &
    $p_{\mathrm{LOS}}(d)$ & LOS probability  \\

    $PL_{\mathrm{LOS/NLOS}}$ & path loss (dB) &
    $p_{\mathrm{cov}}(d)$ & coverage probability \\

    $P_{\mathrm{tx}}^{\mathrm{dBm}}$ & transmit power (dBm) &
    $G_{\mathrm{tx}},G_{\mathrm{rx}}$ & Tx/Rx antenna gains (dB) \\

    $N_{\mathrm{dBm}}$ & receiver noise floor (dBm) &
    $PL_{\max}$ & max.\ tolerable path loss (dB) \\

    $d_{\max}$ & distance at target reliability &
    $\Phi(\cdot)$ & standard normal CDF \\

    $p^\star$ & target coverage/reliability level &
    & \\
    \hline

    $H$ & autoregressive rollout horizon &
    $S_{\mathrm{buf}}$ & buffer size (history) \\

    $L$ & history/window length &
    $D$ & shared latent dimension \\

    $d_{\mathrm{top}}$ & teleoperator force dimension &
    $d_{\mathrm{op}}$ & operator trajectory dimension \\

    $X^{\mathrm{top}}$ & teleoperator input
    ($\mathbb{R}^{L\times d_{\mathrm{top}}}$) &
    $X^{\mathrm{op}}$ & operator trajectory
    ($\mathbb{R}^{L\times d_{\mathrm{op}}}$) \\

    $\mathbf{y}_t,\widehat{\mathbf{y}}_t$ & true / predicted force vector
    ($\mathbb{R}^{d_{\mathrm{top}}}$) &
    $Y^{(H)},\widehat{Y}^{(H)}$ & rollout force sequence
    ($\mathbb{R}^{H\times d_{\mathrm{top}}}$) \\

    $S_{\mathrm{top}},S_{\mathrm{op}}$ & hidden-state sequences &
    $r_{\mathrm{top}},r_{\mathrm{op}}$ & encoder summaries \\

    $h$ & \# attention heads &
    $d_h$ & head dimension ($D{=}h\,d_h$) \\

    $q_i,K_i,V_i$ & query, keys, values (head $i$) &
    $\alpha_i$ & attention weights (head $i$) \\

    $a_i$ & head context (head $i$) &
    $a$ & fused attention context \\

    $W_Q,W_K,W_V$ & projection matrices &
    $W_O$ & output projection \\

    $W_1,W_2,b_1,b_2$ & prediction head parameters &
    $\sigma(\cdot)$ & activation (ReLU) \\

    $\mathcal{L}$ & total training loss &
    $\mathcal{L}_{\mathrm{mse}}$ & MSE loss component \\

    $\mathcal{L}_{\mathrm{rel}}$ & relative error component &
    $\lambda_{\mathrm{mse}},\lambda_{\mathrm{rel}}$ & loss weights \\

    $\tau$ & threshold for relative error &
    $\epsilon_e$ & teacher-forcing probability \\

    $E$ & total epochs &
    $T_{\mathrm{thr}}$ & restoration threshold (force units) \\
    \hline
  \end{tabular}%
  }
\end{table}

% ---------- System Model ----------------------------------------------------------

\section{System Model}
\label{sec:wireless}
We consider a time-varying effective SNR process that incorporates large-scale shadowing, small-scale fading, and temporal correlation. Based on the effective SNR, the bit error rate (BER) is obtained for a given modulation, from which the packet error rate (PER) and goodput can be derived.

\subsection{Temporal SNR Process}
Let $\mu$ denote the average SNR (dB), $\sigma_{\mathrm{sh}}$ the standard deviation of log-normal shadowing~\cite{Szyszkowicz2010}, and $\rho\in[0,1)$ a temporal correlation parameter. The instantaneous shadowing is modeled as:
\begin{equation}
Z_t = S_t + F_t, \qquad S_t \sim \mathcal{N}(0,\sigma_{\mathrm{sh}}^2),
\end{equation}
where $F_t$ represents a small-scale fading component, modeled as Rayleigh fading. Following a temporal correlation model of shadowing from~\cite{Gudmudson1991}, SNR evolves according to a first-order autoregressive model:
\begin{equation}
\mathrm{SNR}_t = \rho \,\mathrm{SNR}_{t-1} + (1-\rho) Z_t .
\end{equation}

\subsection{Forward Error Correction Gain}
Forward error correction (FEC) is incorporated through an effective coding gain $G_{\mathrm{FEC}}(R,\mathrm{SNR})$ in dB, which depends on the code rate $R\in(0,1]$ and the operating SNR region. For analytical clarity, we adopt a piecewise heuristic model in which the gain decreases with higher code rates and saturates at very low or very high SNR. The effective SNR is then given by:
\begin{equation}
\mathrm{SNR}^{\mathrm{eff}}_t = \mathrm{SNR}_t + G_{\mathrm{FEC}}(R,\mathrm{SNR}_t).
\end{equation}

Given the small size of haptic packets, the heuristic model provides adequate accuracy while being considerably more computationally efficient than analytical formulations.

\subsection{Bit Error Rate}
Let $\gamma_t = 10^{\mathrm{SNR}^{\mathrm{eff}}_t/10}$ denote the linear effective SNR. The instantaneous BER is approximated using standard union-bound expressions~\cite{Proakis2007}:
\begin{equation}
\mathrm{BER}(\gamma) = 
\begin{cases}
\dfrac{1}{2}\,\mathrm{erfc}(\sqrt{\gamma}), & \text{BPSK/QPSK},\\[6pt]
0.375\,\mathrm{erfc}(\sqrt{0.4\,\gamma}), & \text{16-QAM}.\\
\end{cases}
\end{equation}

\subsection{Packet Error Rate and Loss Probability}
Assuming independent bit errors, the packet error rate~\cite{Masnikosa2020} for a packet of $N_b$ bits is:
\begin{equation}
\mathrm{PER}_t = 1 - \big(1 - \mathrm{BER}(\gamma_t)\big)^{N_b}.
\label{eq:per}
\end{equation}
The packet success indicator is modeled as a Bernoulli random variable with success probability $1-\mathrm{PER}_t$. The packet loss rate (PLR) over a horizon of $T$ packets is:
\begin{equation}
    \mathrm{PLR} = \frac{N_{\text{LP}}}{T},
\end{equation}
where $N_{LP}$ is the number of lost packets.

\subsection{Spectral Efficiency}
Let $b(\mathsf{M})$ denote the modulation order in bits per symbol and $R$ the code rate. The spectral efficiency is:
\begin{equation}
\eta = b(\mathsf{M}) R .
\end{equation}
Assuming a symbol rate $f_s=\mathcal{B}$, where $\mathcal{B}$ denotes the bandwidth, the coded data rate is:
\begin{equation}
R_{\text{c}} = \eta \mathcal{B}.
\end{equation}
The goodput, i.e., the successfully delivered information rate, is:

\begin{equation}
R_{\mathrm{eff}} = R_{\text{c}} (1 - \mathrm{PER}_t).
\end{equation}
Substituting for $R_{\text{c}}$ yields:
\begin{equation}
R_{\mathrm{eff}} = b(\mathsf{M}) R\,\mathcal{B} \,(1 - \mathrm{PER}_t).
\end{equation}

\subsection{Path Loss Model}

We adopt the 3GPP Urban Macro (UMa) path loss model of 3GPP TR 38 901~\cite{3gpp38901} with distance $d$ (in meters) between base station and user equipment. Given a required SNR $\mathrm{SNR}_{\mathrm{req}}$, the maximum tolerable path loss is
\begin{equation}
  PL_{\max} = P_{\mathrm{tx}}^{\mathrm{dBm}} + G_{\mathrm{tx}} + G_{\mathrm{rx}} - \left(N_{\mathrm{dBm}} + \mathrm{SNR}_{\mathrm{req}}^{\mathrm{dB}}\right),
\end{equation}
with transmit power $P_{\mathrm{tx}}$, antenna gains $G_{\mathrm{tx}},G_{\mathrm{rx}}$, and $N_{\mathrm{dBm}}$ is the receiver noise floor, i.e., the thermal noise over the system bandwidth plus the receiver noise figure.

The coverage probability at distance $d$ is then given by the LOS/NLOS mixture with lognormal shadowing:
\begin{multline}
  p_{\mathrm{cov}}(d) = p_{\mathrm{LOS}}(d)\,
  \Phi\!\left(\tfrac{PL_{\max}-\mathrm{PL}_{\mathrm{LOS}}(d)}{\sigma_{\mathrm{LOS}}}\right) \\
  + \bigl(1-p_{\mathrm{LOS}}(d)\bigr)\,
  \Phi\!\left(\tfrac{PL_{\max}-\mathrm{PL}_{\mathrm{NLOS}}(d)}{\sigma_{\mathrm{NLOS}}}\right),
\end{multline}
where $p_{\mathrm{LOS}}(d)$ is the LOS probability, $\sigma_{\mathrm{LOS}},\sigma_{\mathrm{NLOS}}$ are shadowing standard deviations in~dB, and $\Phi(\cdot)$ is the standard normal CDF. The parameters $\mathrm{PL}_{\mathrm{LOS}}(d)$, $\mathrm{PL}_{\mathrm{NLOS}}(d)$, and $p_{\mathrm{LOS}}(d)$ follow the 3GPP UMa model defined in TR~38.901~\cite{3gpp38901}, with shadowing standard deviations $\sigma_{\mathrm{LOS}}{=}4$~dB and $\sigma_{\mathrm{NLOS}}{=}6$~dB as listed in Table~\ref{tab:sim_channel_params}.

Finally, the maximum coverage distance $d_{\max}$ for a target reliability $p^\star$ is defined as the largest $d$ such that $p_{\mathrm{cov}}(d) \ge p^\star$. This is solved efficiently by bisection, exploiting the monotonic decrease of $p_{\mathrm{cov}}(d)$ with distance.

Under Rayleigh fading, the diversity order $L_{\mathrm{div}}$ represents the number of independent faded copies of a packet (e.g., across frequency or antennas). The linear effective SNR is then:
 \begin{equation}
     \gamma_{\mathrm{eff}} = \frac{1}{L_{\mathrm{div}}}\sum_{i=1}^{L_{\mathrm{div}}} \gamma_i,
     \label{eq:gamma_eff_simple}
 \end{equation}
 with $\gamma_i$ the instantaneous branch SNR\cite{Proakis2007}. Increasing $L_{\mathrm{div}}$ reduces the variance of $\gamma_{\text{eff}}$ by a factor of $1/L_{\mathrm{div}}$, thereby mitigating deep fades and improving reliability through diversity combining. We set this value to $L_{\mathrm{div}} = 3$ channels.

\begin{algorithm}[t]
\caption{Haptic Packet-Loss Restoration (Runtime, absolute $T_{\mathrm{thr}}$)}
\label{alg:plr_restore_abs}
\begin{algorithmic}[1]
\Require history length $L$; restoration threshold $T_{\mathrm{thr}}$; model $f_\theta$; samples $\{x_i\}_{i=1}^N$
\Ensure Effective PLR and restoration rate
\State Buffer $M$ of size $L$; flag: filled $\gets$ False
\State Counters: total $\gets 0$, lost $\gets 0$, restored $\gets 0$
\For{$i=1$ to $N$}
  \State Transmit $x_i$; total $\gets$ total $+1$
  \If{success}
    \State Append $x_i$ to $M$; if $|M|=L$ then filled$\gets$True
  \Else
    \State lost $\gets$ lost $+1$
    \If{filled = False} \State Append zero to $M$; \textbf{continue} \EndIf
    \State $X_i^{\mathrm{top}} \gets$ last $L$ force vectors from $M$
    \State Obtain $X_i^{\mathrm{op}}$ from the operator-command buffer
    \State $\hat{x}_i \gets f_\theta(X_i^{\mathrm{top}},X_i^{\mathrm{op}})$
    \State $e \gets \frac{1}{3}\sum_{c=1}^{3}\big|\hat{x}_{i,c}-x_{i,c}\big|$
    \If{$e \le T_{\mathrm{thr}}$} \State restored $\gets$ restored $+1$; append $\hat{x}_i$ \Else \State append zero \EndIf
  \EndIf
\EndFor
\State effective\_plr $= (\text{lost}-\text{restored})/\max(1,\text{total})$
\State restoration\_rate $= \begin{cases}\text{restored}/\text{lost} & \text{if lost}>0\\ 0 & \text{otherwise}\end{cases}$
\State \Return effective\_plr, restoration\_rate
\end{algorithmic}
\end{algorithm}

\subsection{Haptic Packet Restoration Pipeline}
\rev{To bridge the wireless channel error model with the application-layer control loop, we implement a runtime restoration pipeline at the receiver. When a packet is successfully decoded, it is appended to the local history buffer. When a communication outage occurs (i.e., the packet fails the error check defined in Eq.~\eqref{eq:per}), the system invokes the predictive model $f_\theta$ to estimate the lost haptic data.}

\rev{To ensure safety and stability in the teleoperation loop, the estimated force is evaluated against an absolute error threshold $T_{\mathrm{thr}}$. If the estimation error remains within this threshold, the packet is considered to have been successfully restored. Otherwise, it is recorded as a definitive loss. This end-to-end procedure for calculating the effective Packet Loss Rate (PLR) and the restoration rate is formally detailed in Algorithm~\ref{alg:plr_restore_abs}.}

% Change \Require and \Ensure keywords in algpseudocode
\algrenewcommand\algorithmicrequire{\textbf{Input:}}
\algrenewcommand\algorithmicensure{\textbf{Output:}}

%------------------ xHAP -----------------------------------------------

\begin{figure*}[!t]
 \centering
 \includegraphics[width=1.0\textwidth]{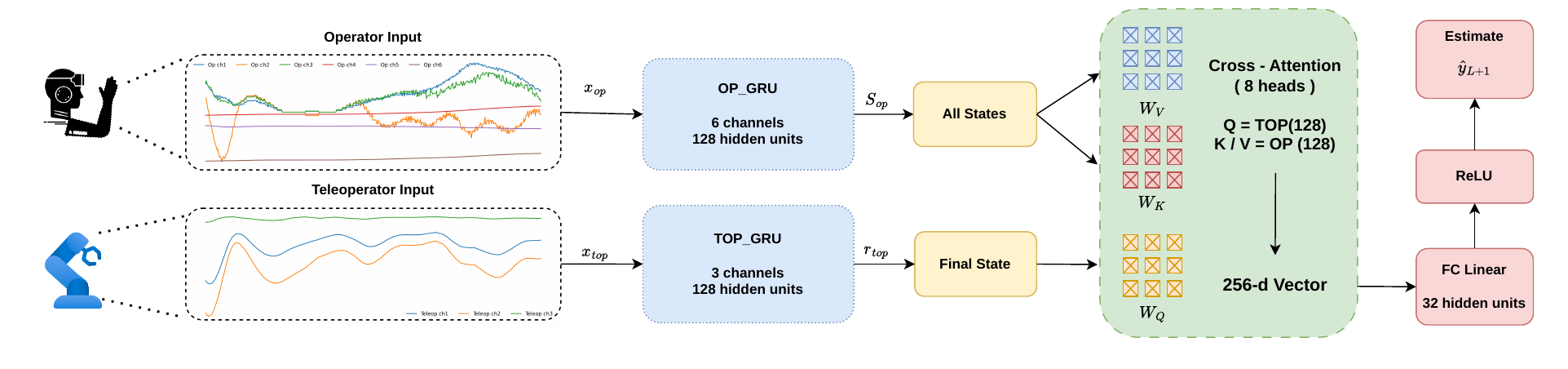}
\caption{\rev{xHAP cross-attention architecture. Separate GRU encoders process the teleoperator force history and operator trajectory inputs. The final teleoperator hidden state is used as the query, while the operator hidden-state sequence provides the keys and values for multi-head cross-attention. The resulting attention context is concatenated with the teleoperator representation and passed through a feed-forward prediction head to produce the next-step force estimate $\widehat{\mathbf{y}}_{L+1}$.}}
 \label{fig:xHAP_architecture}
\end{figure*}

\section{xHAP: Cross-Attention Haptic Estimator}
\label{sec:estimator}

In this section, we describe the cross-modal attention-based estimator for \rev{next-step haptic force prediction and autoregressive multi-step restoration}. As shown in Fig.~\ref{fig:xHAP_architecture}, the model is designed to exploit both historical teleoperator feedback and operator trajectory information, enabling robust prediction under packet loss conditions. \rev{Algorithm~\ref{alg:xhap_estimator} summarizes the encoding, cross-attention, fusion, and prediction steps.}

\subsection{Problem Setup}
Let \(L\) denote the teleoperator history length. For each training window (batch dimension omitted), the model uses two temporal input sequences and one next-step target:
\begin{itemize}
    \item Teleoperator history: \(X^{\mathrm{top}} \in \mathbb{R}^{L \times d_{\mathrm{top}}}\), representing the most recent sequence of 3D forces.
    \item Operator trajectory: \(X^{\mathrm{op}} \in \mathbb{R}^{L \times d_{\mathrm{op}}}\), including the operator’s 3D position and velocity over the input window. Unlike teleoperator forces, which are predicted autoregressively, operator states are received continuously during inference, ensuring access to up-to-date observations at every timestep.
    \item Ground-truth teleoperator force: \(Y \in \mathbb{R}^{d_{\mathrm{top}}}\), serving as the prediction target.
\end{itemize}

During both training and inference, the operator trajectory is treated as an up-to-date input stream. The estimator never accesses current teleoperator forces \(Y\). We stack time steps as vector columns:
\begin{equation}
    X^{\mathrm{top}} = \big[x^{\mathrm{top}}_{1},\dots,x^{\mathrm{top}}_{L}\big]^T, 
    \qquad x^{\mathrm{top}}_{t} \in \mathbb{R}^{d_{\mathrm{top}}},
\end{equation}
\begin{equation}
    X^{\mathrm{op}} = \big[x^{\mathrm{op}}_{1},\dots,x^{\mathrm{op}}_{L}\big]^T, 
    \qquad x^{\mathrm{op}}_{t} \in \mathbb{R}^{d_{\mathrm{op}}},
\end{equation}
\begin{equation}
    Y = \mathbf{y}_{L+1} \in \mathbb{R}^{d_{\mathrm{top}}}.
\end{equation}
We set \(d_{\mathrm{top}}=3\) (3D force) and \(d_{\mathrm{op}}=6\) (3D position and 3D velocity). The learning objective is:
\begin{equation}
    f_\theta: \big(X^{\mathrm{top}}, X^{\mathrm{op}}\big) \mapsto \widehat{\mathbf{y}}_{L+1}\in \mathbb{R}^{d_{\mathrm{top}}},
\end{equation}
where $\widehat{\mathbf{y}}_{L+1}$ denotes the estimated next-step force vector.

\subsection{Modality-Specific Encoders}
We employ two GRU-based encoders that project each modality with the same embedding dimensionality $D = 128$. We use consistent notation \(E_{\mathrm{top}}=\mathrm{GRU}_{\mathrm{top}}\) and \(E_{\mathrm{op}}=\mathrm{GRU}_{\mathrm{op}}\):
\begin{equation}
    E_{\mathrm{top}}:\ \mathbb{R}^{L \times d_{\mathrm{top}}}\to\mathbb{R}^{L \times D}.
\end{equation}
\begin{equation}
    E_{\mathrm{op}}:\ \mathbb{R}^{L \times d_{\mathrm{op}}}\to\mathbb{R}^{L \times D}.
\end{equation}
Applying the encoders yields hidden-state sequences:
\begin{equation}
    S_{\mathrm{top}} = E_{\mathrm{top}}(X^{\mathrm{top}}) = \{h^{\mathrm{top}}_t\}_{t=1}^{L}\in\mathbb{R}^{L\times D}.
\end{equation}
\begin{equation}
    S_{\mathrm{op}} = E_{\mathrm{op}}(X^{\mathrm{op}}) = \{h^{\mathrm{op}}_t\}_{t=1}^{L}\in\mathbb{R}^{L\times D}.
\end{equation}

\subsection{Cross-Attention Fusion}

\rev{Unlike early fusion~\cite{baltrusaitis2019multimodal}, which combines modalities at the raw-feature or feature-representation level, or late fusion~\cite{baltrusaitis2019multimodal}, which combines modality-specific representations or decisions after separate processing and may miss fine-grained cross-modal temporal interactions, cross-attention provides dynamic, input-dependent temporal alignment. Compared with static concatenation, cross-attention is also more selective, since it does not combine the two streams through a fixed content-independent representation. By allowing each teleoperator query to evaluate all operator timesteps, cross-attention dynamically weights the most relevant moments to handle the variable lag between an action and the resulting force. Our ablation study in Section~\ref{sec:comparison} (Table~\ref{tab:ablation}) highlights the importance of this selectivity.}

To integrate both modalities, we employ multi-head scaled dot-product attention~\cite{Vaswani2017}. The teleoperator embedding $r_{\mathrm{top}} = h_L^{\mathrm{top}}$ serves as a single \emph{query}. We use the final hidden state of the encoding as the query, as it is the most informative state for the teleoperator sequence. The sequence \(S_{\mathrm{op}}=\{h^{\mathrm{op}}_t\}_{t=1}^{L}\) provides \emph{keys} and \emph{values}.

For each head \(i\in\{1,\dots,h\}\) with \(d_h=D/h\):
\begin{equation}
    \label{eq:query}
    q_i = W_Q^{(i)} r_{\mathrm{top}} \in \mathbb{R}^{d_h},
\end{equation}
\begin{equation}
    \label{eq:key}
    K_i = S_{\mathrm{op}} W_K^{(i)} \in \mathbb{R}^{L\times d_h},
\end{equation}
\begin{equation}
    \label{eq:value}
    V_i = S_{\mathrm{op}} W_V^{(i)} \in \mathbb{R}^{L\times d_h},
\end{equation}
The attention is then calculated as follows:
\begin{equation}
    \label{eq:attn}
    \alpha_i = \mathrm{softmax}\!\left(\tfrac{K_i q_i}{\sqrt{d_h}}\right) \in \mathbb{R}^{L},
\end{equation}
\begin{equation}
    \label{eq:attn2}
    a_i = V_i^T \alpha_i \in \mathbb{R}^{d_h},
\end{equation}
\begin{equation}
    a = W_O [a_1; \dots; a_h] \in \mathbb{R}^{D}.
\end{equation}
Finally, the attention context is combined with the teleoperation output by concatenation:
\begin{equation}
    z = [r_{\mathrm{top}}; a] \in \mathbb{R}^{2D}.
\end{equation}
This design enables the teleoperator embedding to attend to temporally relevant operator states, improving predictive robustness.
\begin{algorithm}[t]

\caption{Cross-Attention GRU Estimator}
\label{alg:xhap_estimator}
\begin{algorithmic}[1]
\Require Teleoperator history $X^{\mathrm{top}}\in\mathbb{R}^{L\times d_{\mathrm{top}}}$; 
         operator trajectory $X^{\mathrm{op}}\in\mathbb{R}^{L\times d_{\mathrm{op}}}$; 
         number of heads $h$, head dimension $d_h$, latent $D{=}h\,d_h$
\Ensure Predicted force vector $\widehat{\mathbf{y}}_{L+1}\in\mathbb{R}^{d_{\mathrm{top}}}$

\State Encode teleoperator history: $S_{\mathrm{top}} \gets \mathrm{GRU}_{\mathrm{top}}(X^{\mathrm{top}})$
\State Encode operator trajectory: $S_{\mathrm{op}} \gets \mathrm{GRU}_{\mathrm{op}}(X^{\mathrm{op}})$
\State $r_{\mathrm{top}} \gets S_{\mathrm{top}}[L]$

\For {Each attention head $i \in \{1,\dots,h\}$}
  \State Compute projections as in Eqs.~\eqref{eq:query}--\eqref{eq:value}
  \State Compute attention score as in Eqs.~\eqref{eq:attn},\eqref{eq:attn2}
\EndFor

\State Concatenate heads: $a \gets W_O [a_1;\dots;a_h] \in \mathbb{R}^{D}$
\State Fuse modalities: $z \gets [r_{\mathrm{top}}; a] \in \mathbb{R}^{2D}$
\State Predict: $\widehat{\mathbf{y}}_{L+1} \gets W_2\,\sigma(W_1 z + b_1) + b_2$
\State \Return $\widehat{\mathbf{y}}_{L+1}$
\end{algorithmic}
\end{algorithm}
\subsection{Prediction Head}
The fused representation \(z\) is mapped to the next-step force estimate using a two-layer feed-forward network:
\begin{equation}
    \widehat{\mathbf{y}}_{L+1} =  W_2 \,\sigma(W_1 z + b_1) + b_2 \in \mathbb{R}^{d_{\mathrm{top}}},
\end{equation}
where \(\sigma(\cdot)\) denotes the ReLU activation. This design keeps the estimator lightweight while preserving predictive capacity. The complete step-by-step procedure for encoding the modalities, computing the cross-attention, and generating the final force estimate is summarized in Algorithm~\ref{alg:xhap_estimator}.

\subsection{Training Objective}
\rev{Although xHAP outputs a single next-step force estimate, training is performed over an autoregressive rollout horizon. Specifically, the estimator is rolled out for $H$ steps using scheduled teacher forcing, and the loss is accumulated over the resulting force sequence.} We optimize the estimator using a composite loss that combines the MSE and relative error, \rev{enforcing both} absolute accuracy and robustness. This combination prevents bias toward high-magnitude forces and maintains stable performance across varying contact dynamics common in haptic interactions:

\begin{equation}
    \mathcal{L} = \lambda_{\mathrm{mse}} \,\mathcal{L}_{\mathrm{mse}} + \lambda_{\mathrm{rel}} \,\mathcal{L}_{\mathrm{rel}},
\end{equation}
\begin{equation}
    \mathcal{L}_{\mathrm{mse}} = \frac{1}{d_{\mathrm{top}}H}
    \sum_{t=1}^H \sum_{c=1}^{d_{\mathrm{top}}}
    \big(\widehat{Y}^{(H)}_{t,c} - Y^{(H)}_{t,c}\big)^2,
\end{equation}
\begin{equation}
\begin{aligned}
    \mathcal{L}_{\mathrm{rel}}
    &= \frac{1}{|\mathcal{S}|}
    \sum_{(t,c)\in \mathcal{S}}
    \frac{\big|\widehat{Y}^{(H)}_{t,c} - Y^{(H)}_{t,c}\big|}
    {|Y^{(H)}_{t,c}|}, \\
    \mathcal{S}
    &= \{(t,c): |Y^{(H)}_{t,c}| > \tau\}.
\end{aligned}
\end{equation}

where $\tau$ denotes a threshold value used to exclude negligible force values from $\mathcal{L}_{\mathrm{rel}}$, preventing instability due to inflated relative errors. We use loss weights \(\lambda_{\mathrm{mse}} = \lambda_{\mathrm{rel}} = 0.5\) and \(\tau=0.01\).

\subsection{Autoregressive Output}
Although the estimator is trained on fixed-length windows, it can be applied autoregressively. After predicting \(\widehat{\mathbf{y}}_{L+1}\), the estimate is fed back into the input window to obtain $(\widehat{\mathbf{y}}_{L+2}, \widehat{\mathbf{y}}_{L+3}, \dots,\widehat{\mathbf{y}}_{L+H})$, where $H$ is the autoregressive rollout horizon. In this sliding-window approach, predicted forces are combined with the true operator commands at each step. The setup maintains force continuity under consecutive packet losses, while periodic reception of ground-truth packets re-calibrates the model and prevents unbounded error growth, supporting robust wireless teleoperation.

%------------------------- Model Performance evaluation -----------------------------------------------

\section{\rev{Haptic Force Estimation Performance}}
\label{sec:comparison}
This section provides evaluation results for the proposed xHAP cross-attention model for reliable haptic signal estimation and its implications for wireless channel performance. We describe the training setup, compare xHAP with competing architectures across multiple teleoperation activities, and analyze both quantitative metrics and feature-level insights to assess accuracy, efficiency, and generalization capability.

\subsection{Model and Training Setup}

For our model configuration, we set the encoder dimensionality to $D = 128$ and use $h = 8$ attention heads, yielding a per-head dimension of
\[
d_h = \frac{D}{h} = \frac{128}{8} = 16.
\]
Since the fused representation has size $2D = 2 \times 128 = 256$, the two-layer feed-forward network has parameters:
\[
W_1 \in \mathbb{R}^{32 \times 256}, \qquad b_1 \in \mathbb{R}^{32},
\]
followed by:
\[
W_2 \in \mathbb{R}^{3 \times 32}, \qquad b_2 \in \mathbb{R}^{3}.
\]

Training uses Adam optimizer with a step learning-rate (LR) scheduler, with a step size $\mathrm{SLR} = 10~\mathrm{epochs}$, and the gamma $\gamma_\mathrm{LR}=0.5$.  Moreover, we adopt Scheduled Teacher Forcing~\cite{LambTforcing}, a common strategy in sequence prediction tasks where the probability of feeding the ground-truth force value instead of the model’s previous output decreases over training epochs. 
Let $\epsilon_e$ denote the teacher-forcing probability at epoch $e$. At each prediction step, the true previous force is used with probability $\epsilon_e$, and the model’s estimate otherwise. The schedule follows a linear decay given by
\begin{equation}
\epsilon_e = 1 - \frac{e}{E},
\end{equation}
where $E$ is the total number of training epochs.

We use haptic traces collected with a Phantom Omni device for the five activities mentioned in Table~\ref{tab:model_performance}: Dynamic Object Pushing (Dyn. Push), Dynamic Object Tapping (Dyn. Tap), Rigid Body Interaction (RB Int), Rigid Body Push and Hold (RB P\&H), and Rigid Body Tapping (RB Tap). Each activity on the original dataset spans 120s and is sampled at 1 kHz, yielding $120{,}000$ samples per task~\cite{Daniel_traces}. Since the haptic devices must be activated at the start and deactivated at the end of each task, we discard the first and last $10{,}000$ samples of each trace to remove activation and shutdown artifacts, resulting in $100{,}000$ samples per task. Training and validation are performed on separate repetitions of the same activities, and estimator performance is reported using only the validation error. \rev{The separate repetitions, together with the inserted delay used during testing, introduce small oscillations and meaningful differences between training and validation dynamics.}

\begin{table}[!htbp]
  \caption{\rev{Component and encoder ablation. $10$-step validation MSE. Each configuration independently tuned, spanning $N\!=\!30$ with median pruning. $\Delta$ is the MSE increase relative to the full xHAP model.}}
  \label{tab:ablation}
  \centering
  \renewcommand{\arraystretch}{1.1}
  \resizebox{\columnwidth}{!}{%
  \begin{tabular}{lccccc}
    \toprule
    \textbf{\rev{Configuration}} & \textbf{\rev{Params (K)}} & \textbf{\rev{GPU (MB)}} & \textbf{\rev{MSE}} $\downarrow$ & \textbf{\rev{$\Delta$ MSE}} & \textbf{\rev{$\Delta$ (\%)}} \\
    \midrule
    \multicolumn{6}{l}{\textit{\rev{Component ablation (leave-one-out)}}} \\
    \textbf{\rev{xHAP (full)}}        & \rev{178} & \rev{8.8} & \textbf{\rev{0.0042}} & \rev{---} & \rev{---} \\
    \textbf{\rev{No Cross-attention}} & \rev{112} & \rev{7.6} & \rev{0.0105} & \rev{$+0.0063$} & \rev{$+149.0\%$} \\
    \textbf{\rev{No OP-branch}}     & \rev{60}  & \rev{6.2} & \rev{0.0056} & \rev{$+0.0014$} & \rev{$+33.3\%$} \\
    \textbf{\rev{No Closed-loop}}     & \rev{178} & \rev{8.8} & \rev{0.0110} & \rev{$+0.0068$} & \rev{$+162.0\%$} \\
    \midrule
    \multicolumn{6}{l}{\textit{\rev{Recurrent encoder (matched settings)}}} \\
    \textbf{\rev{GRU (xHAP)}} & \rev{178} & \rev{8.8}  & \textbf{\rev{0.0042}} & \rev{---} & \rev{---} \\
    \textbf{\rev{LSTM}}       & \rev{232} & \rev{10.4} & \rev{0.0122} & \rev{$+0.0080$} & \rev{$+190.5\%$} \\
    \textbf{\rev{RNN}}        & \rev{121} & \rev{6.9}  & \rev{0.0732} & \rev{$+0.0690$} & \rev{$+1642.9\%$} \\
    \bottomrule
  \end{tabular}%
  }
\end{table}

\begin{table}[!htbp]
  \caption{\rev{Scale-hyperparameter selection via partial dependence over the $8$ completed (non-pruned) trials of the Bayesian search (remaining hyperparameters marginalized). Adopted values in bold.}}
  \label{tab:hyperparams}
  \centering
  \renewcommand{\arraystretch}{1.1}
  \resizebox{\columnwidth}{!}{%
  \begin{tabular}{llccc}
    \toprule
    \textbf{\rev{Hyperparameter}} & \textbf{\rev{Value}} & \textbf{\rev{Trials}} & \textbf{\rev{Best MSE}} $\downarrow$ & \textbf{\rev{Median MSE}} \\
    \midrule
    \rev{Hidden size}     & \rev{64}           & \rev{1} & \rev{0.0212}          & \rev{0.0212} \\
                    & \textbf{\rev{128}} & \rev{7} & \textbf{\rev{0.0042}} & \textbf{\rev{0.0053}} \\
    \addlinespace
    \rev{Attention heads} & \rev{4}            & \rev{3} & \rev{0.0046}          & \rev{0.0061} \\
                    & \textbf{\rev{8}}   & \rev{5} & \textbf{\rev{0.0042}} & \textbf{\rev{0.0053}} \\
    \addlinespace
    \rev{FC width}        & \textbf{\rev{32}}  & \rev{7} & \textbf{\rev{0.0042}} & \textbf{\rev{0.0053}} \\
                    & \rev{64}           & \rev{1} & \rev{0.0212}          & \rev{0.0212} \\
    \bottomrule
  \end{tabular}%
  }
\end{table}

\subsection{\rev{Component and Design Ablation}}
\label{subsec:ablation}
\rev{To identify which design choices drive xHAP's accuracy, we perform a controlled leave-one-out ablation: starting from the full model, each variant removes a single design element while leaving the rest intact. Each variant is independently tuned by Bayesian optimization, spanning $N=30$ trials with median pruning, and scored by the same validation MSE. The results are summarized in Table~\ref{tab:ablation}.}

\begin{itemize}
    \item \rev{\textbf{Cross-attention:} Replacing the attention fusion with a static, content-independent concatenation $[\mathbf{h}^{\mathrm{top}}_{L}\Vert\mathbf{h}^{\mathrm{op}}_{L}]$ raises the MSE from $0.0042$ to $0.0105$ ($+149\%$). This confirms the performance gain comes from the dynamic attention mechanism absorbing variable lead-lag, rather than merely the presence of operator data.}
    
    \item \rev{\textbf{Operator branch:} Including the operator-command branch allows xHAP to exploit the relationship between operator motion and teleoperator force feedback. Removing this branch raises the MSE to $0.0056$ ($+33\%$), confirming that operator-side motion cues contribute directly to the restoration accuracy.}
    
    \item \rev{\textbf{Closed-loop training:} Training an open-loop variant (fixing $\epsilon_e\!=\!1$) induces exposure bias, as the training input distribution differs from the autoregressive rollout at deployment. This raises the MSE to $0.0110$ ($+162\%$), the largest single error contribution.}
\end{itemize}

\rev{Comparing recurrent encoders under matched settings, the GRU attains the lowest MSE ($0.0042$) against the LSTM ($0.0122$) and a RNN ($0.0732$). Scale hyperparameters were selected via Bayesian search partial dependence (Table~\ref{tab:hyperparams}), aligning with the cross-architecture findings in Table~\ref{tab:model_performance}.}

\begin{table}[!t]
  \caption{Deep Neural Network Model Performance}
  \label{tab:model_performance}
  \centering
  \renewcommand{\arraystretch}{1.15}
  \resizebox{\columnwidth}{!}{%
  \begin{tabular}{lccc}
   \toprule
    & \textbf{xHAP} & \textbf{TCN} & \textbf{ConvLSTMTran} \\
    \hline
    \textbf{Params (K)} & \textbf{178} & 213 & 716 \\
    \textbf{Inference Time (ms)}  & \textbf{0.562} & 1.047 & 1.478 \\
    \textbf{GPU (MB)}   & \textbf{8.8} & 8.9 & 10.9 \\
    \textbf{RAM (MB)}   & \textbf{783.2} & 875.2 & 904.0 \\
    \hline\hline

    \multicolumn{4}{l}{\textbf{Restoration @0.05N (\%)}} \\
    Dyn. Push & \textbf{98.9} & 45.8 & 43.3 \\
    Dyn. Tap & \textbf{99.6} & 65.8 & 73.5 \\
    RB Inter. & \textbf{92.3} & 55.5 & 2.7 \\
    RB P\&H & \textbf{100.0} & 1.7 & 1.6 \\
    RB Tap & \textbf{99.7} & 75.1 & 79.3 \\
    \textbf{Average} & \textbf{97.4} & 48.8 & 40.1 \\
    \hline

    \multicolumn{4}{l}{\textbf{Restoration @0.1N (\%)}} \\
    Dyn. Push & \textbf{99.8} & 62.6 & 45.7 \\
    Dyn. Tap & \textbf{99.8} & 83.0 & 73.9 \\
    RB Inter. & \textbf{100.0} & 87.4 & 3.8 \\
    RB P\&H & \textbf{100.0} & 57.3 & 1.6 \\
    RB Tap & \textbf{99.9} & 84.4 & 79.6 \\
    \textbf{Average} & \textbf{99.9} & 74.9 & 40.9 \\
    \hline\hline

    \multicolumn{4}{l}{\textbf{Relative Restoration @10\% (\%)}} \\
    Dyn. Push & \textbf{99.6} & 85.7 & 13.2 \\
    Dyn. Tap & \textbf{98.1} & 70.3 & 7.2 \\
    RB Inter. & \textbf{99.9} & 95.3 & 22.4 \\
    RB P\&H & \textbf{100.0} & \textbf{100.0} & 0.6 \\
    RB Tap & \textbf{98.9} & 67.9 & 29.9 \\
    \textbf{Average} & \textbf{99.3} & 83.8 & 14.6 \\
    \hline

    \multicolumn{4}{l}{\textbf{Relative Restoration @20\% (\%)}} \\
    Dyn. Push & \textbf{99.8} & 95.3 & 41.1 \\
    Dyn. Tap & \textbf{98.6} & 83.7 & 60.3 \\
    RB Inter. & \textbf{99.9} & 98.7 & 83.0 \\
    RB P\&H & \textbf{100.0} & \textbf{100.0} & 0.8 \\
    RB Tap & \textbf{99.2} & 86.1 & 73.4 \\
    \textbf{Average} & \textbf{99.5} & 92.7 & 51.7 \\
   \bottomrule
  \end{tabular}%
  }
\end{table}

\subsection{xHAP performance evaluation}

We compare our proposed method, xHAP, to a temporal-convolution network (TCN) inspired from~\cite{borovykh2018} and the convolution-LSTM-Transformer model from~\cite{kokkinis2025}.

Table~\ref{tab:model_performance} summarizes model size, inference speed, and restoration accuracy. 
Compared to ConvLSTMTran, xHAP achieves up to $2.63\times$ faster inference and \rev{requires $75\%$ fewer parameters}, while also reducing GPU and RAM usage by up to $19.3\%$ and $13.4\%$, respectively.

At restoration thresholds of $0.05$ and $0.1$, xHAP reconstructs the largest fraction of missing packets, achieving mean restoration rates of $97.4\%$ and $99.9\%$, respectively, whereas competing methods degrade significantly for specific activities.

\begin{figure}[!t]
 \centering
 \includegraphics[width=1\columnwidth]{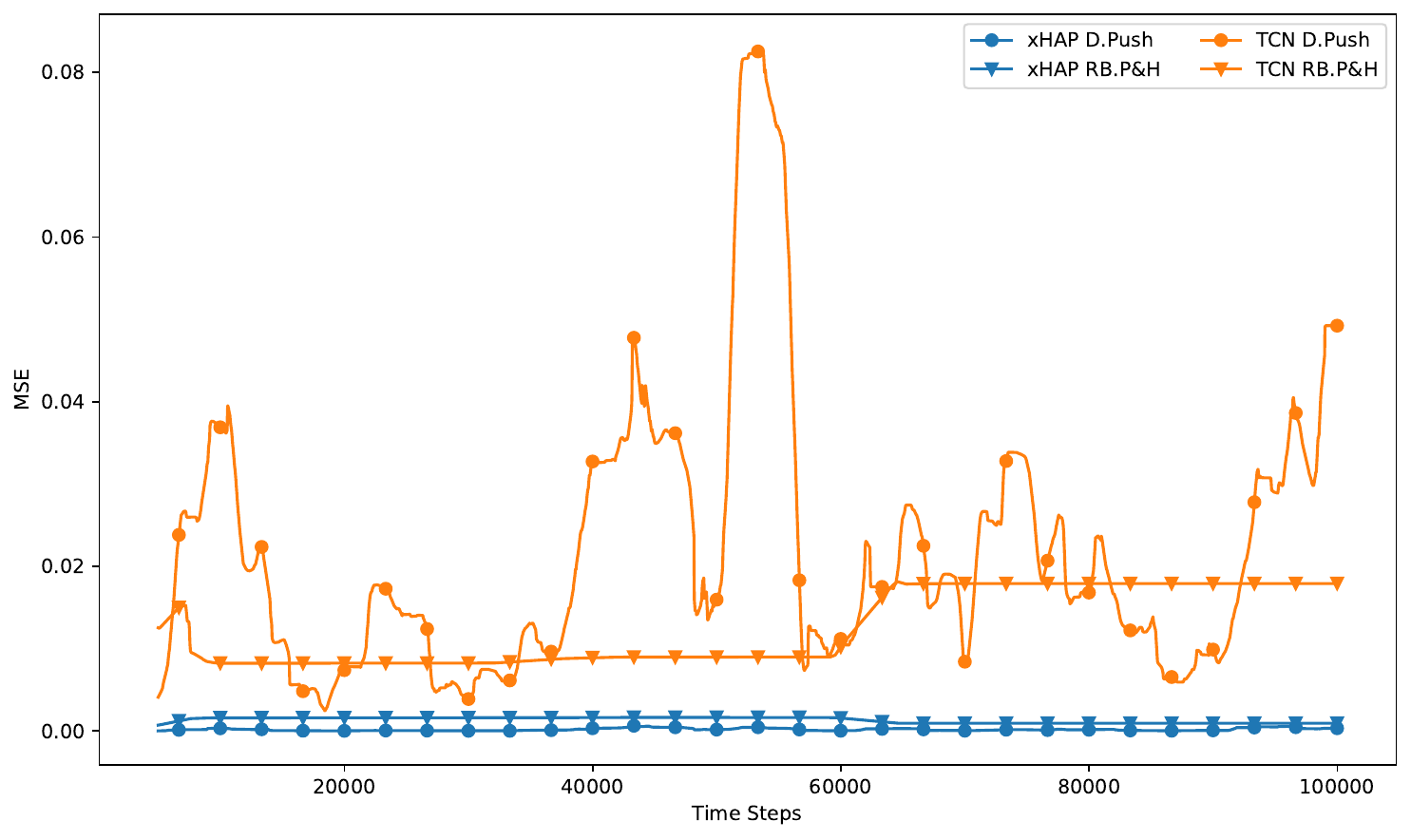}
     \caption{MSE with rolling window over Dyn.Push and RB P\&H.}
 \vspace{-1em}
\label{fig:rolling_mse}
\end{figure}

\begin{figure}[!t]
 \centering
 \includegraphics[width=1\columnwidth]{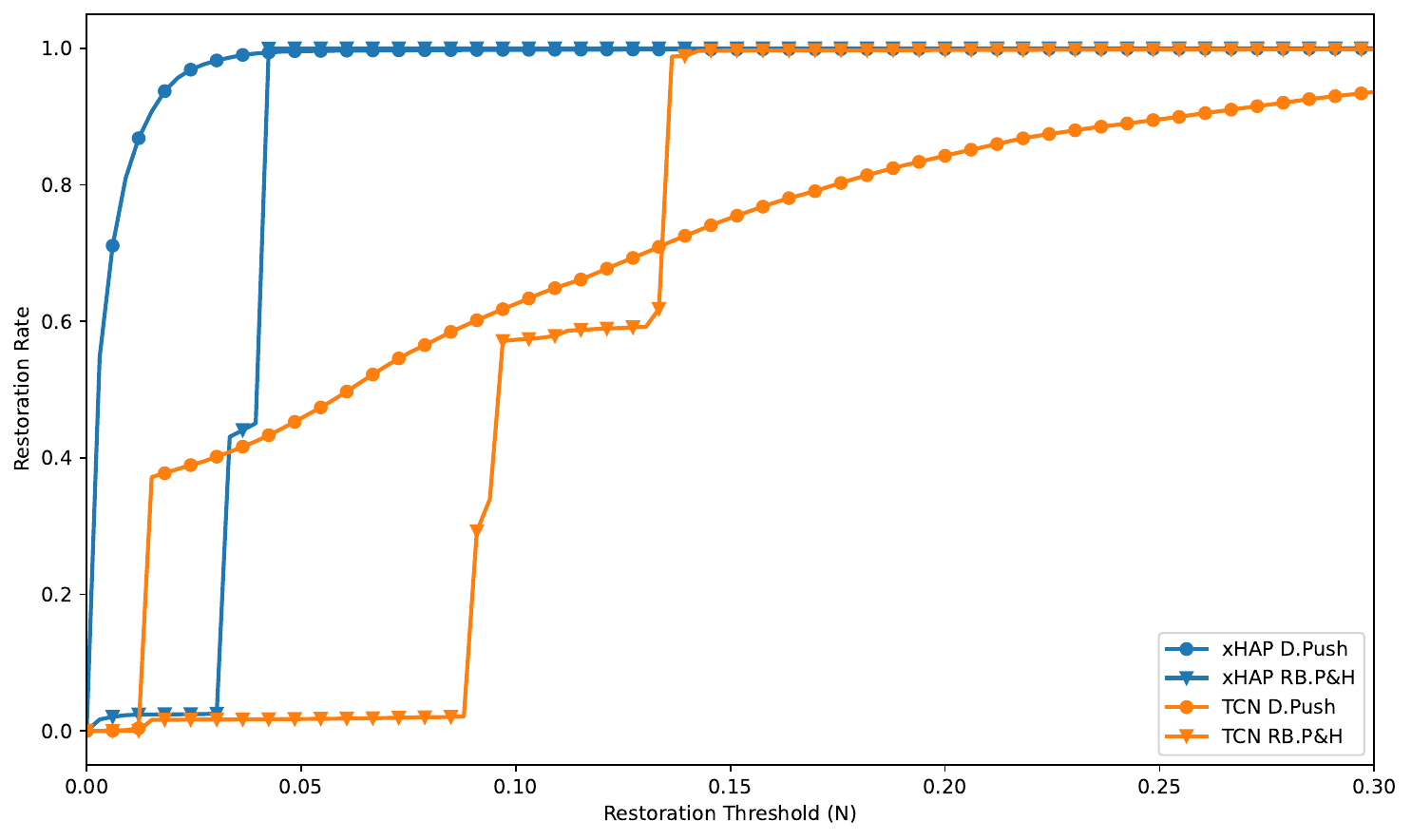}
     \caption{Restoration rate of Dyn.Push and RB P\&H over increasing restoration error threshold.}
 \vspace{-1em}
\label{fig:restoration_rate_vs_threshold}
\end{figure}

Although some activities, such as Dyn.~Tap and RB~Tap, can be reconstructed by more than $70\%$ by all models, Dyn.~Push and RB~Inter. remain more challenging. For these two activities, at a restoration threshold of $0.05$, our method improves restoration performance by $53.1\%$ and $36.8\%$, respectively, compared to the TCN. We visually show the improvement in Fig.~\ref{fig:rolling_mse} and Fig.~\ref{fig:restoration_rate_vs_threshold}. In Fig.~\ref{fig:rolling_mse} we show that \rev{for a rolling average MSE with a window size of 5000 steps}, the xHAP exhibits much lower error compared to the TCN, which fluctuates throughout the activities. The TCN shows a rolling average peak error higher than $0.08$, whereas xHAP peaks at less than 0.01. Similarly, we show in Fig.~\ref{fig:restoration_rate_vs_threshold} that the restoration rate over an increasing restoration threshold is also significantly improved with the proposed method, with the rate being close to $100\%$ near the 0.1 threshold.

For $10\%$ and $20\%$ relative-error thresholds, xHAP consistently achieves the highest restoration accuracy across all tasks, improving performance by up to $85\%$ compared to baselines. 
Dynamic tasks yield slightly higher relative errors, while rigid-body interactions exhibit larger absolute errors.

\rev{From a perceptual perspective, these relative-error results also link the numerical gains to user experience. Human force discrimination is commonly characterized using just-noticeable-difference (JND) or Weber-fraction thresholds, with prior studies reporting force-discrimination thresholds on the order of $7$--$13\%$ depending on the limb, task, and experimental protocol~\cite{Pang1991ForceJND,Feyzabadi2013ForceDiscrimination}. xHAP achieves an average relative restoration rate of $99.3\%$ under a $10\%$ error threshold and remains above $98.1\%$ across all tasks, indicating that most restored samples fall within a perceptually relevant error range. Although this does not replace a dedicated subjective user study, it supports the practical relevance of the observed MSE and restoration-rate improvements.}

\begin{figure}[!t]
 \centering
 \includegraphics[width=1\columnwidth]{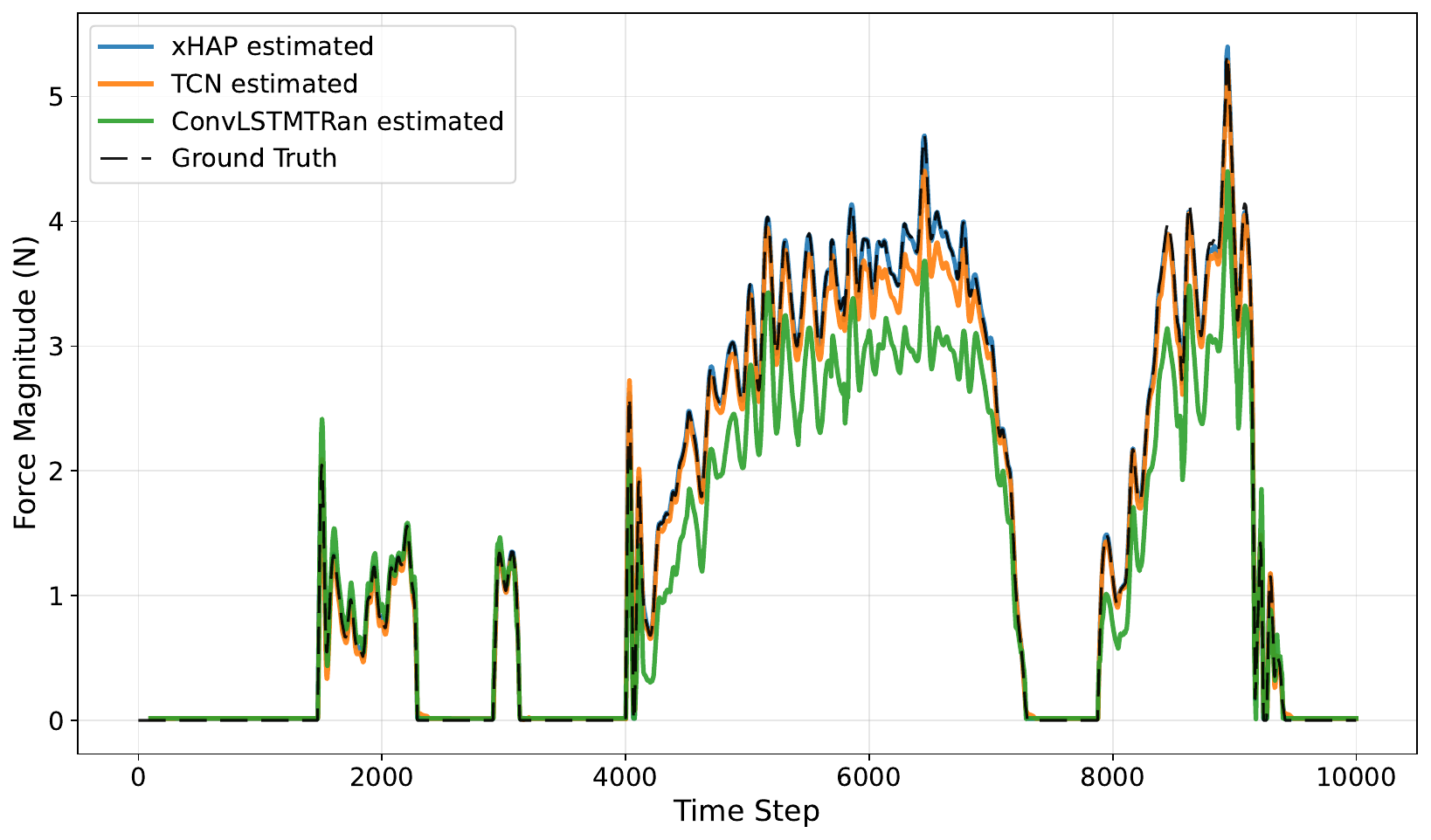}
 \caption{\rev{Comparison of single-step model estimates with the ground truth for the dynamic pushing trace.}}
 \vspace{-1em}
 \label{fig:Estimated_vs_truth}
\end{figure}

\begin{figure}[!t]
 \centering
 \includegraphics[width=1\columnwidth]{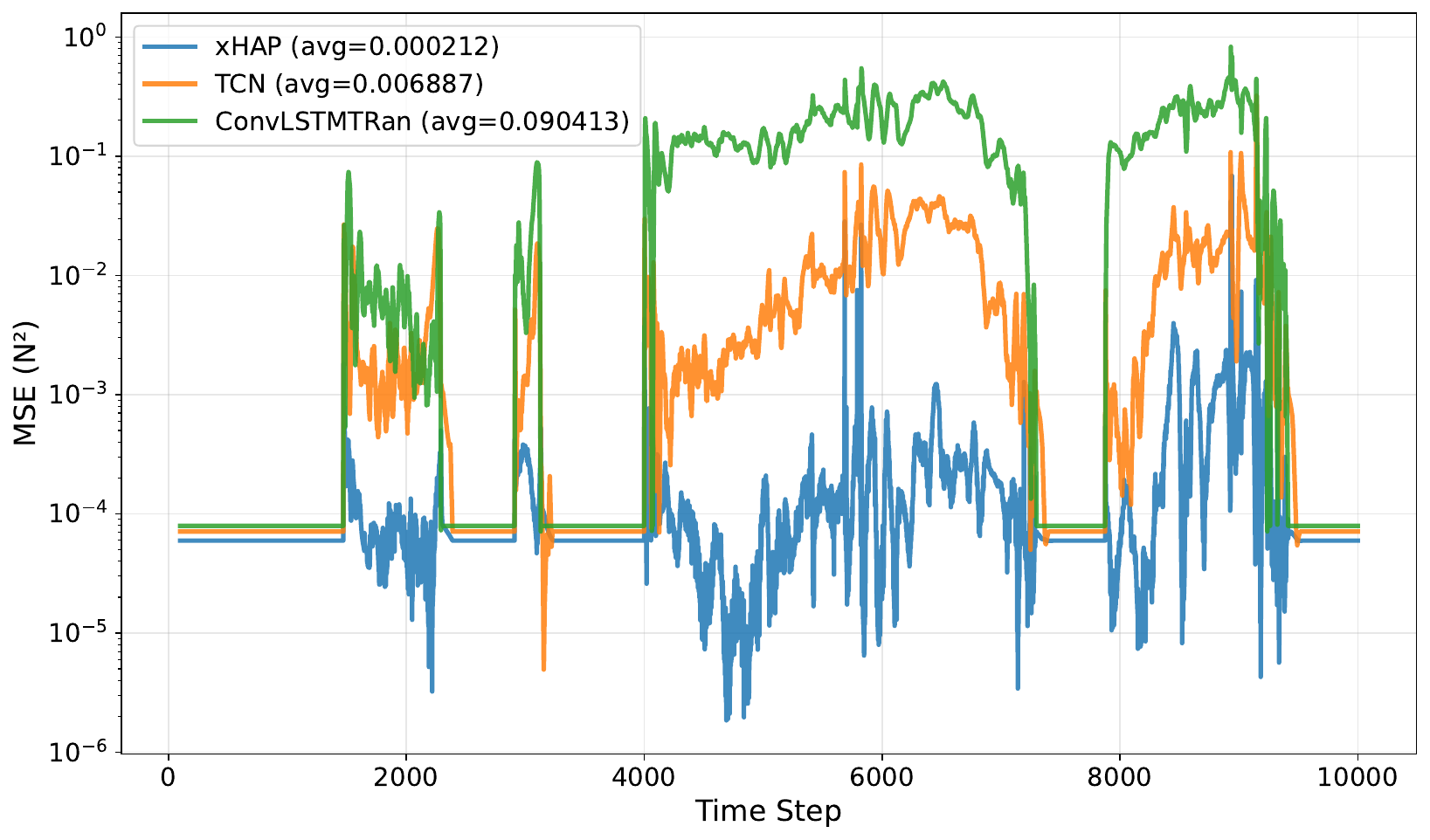}
 \caption{Mean squared error across all models for the dynamic pushing trace.}
 \vspace{-1em}
 \label{fig:MSE_models}
\end{figure}

 In Fig.~\ref{fig:Estimated_vs_truth}, the estimated values from all 3 models are plotted against the magnitude of force of a ground truth trace from the dynamic pushing activity. It is visibly apparent that the proposed method \rev{approximates the ground truth better than the others}, with some regions of the trace deviating slightly from the ground truth. Although the estimates from all models appear visually similar in Fig.~\ref{fig:Estimated_vs_truth}, this visual closeness can be misleading, as subtle deviations may correspond to significant quantitative differences in performance. To further quantify the performance of the model, we plot the MSE for the estimated haptic data trace of the dynamic pushing task in Fig.~\ref{fig:MSE_models}. It becomes apparent that xHAP and TCN perform significantly better than the ConvLSTMTran model. For instance, xHAP achieves the lowest MSE of $2.12\times10^{-4}$.

\begin{figure}[!t]
 \centering
 \includegraphics[width=1\columnwidth]{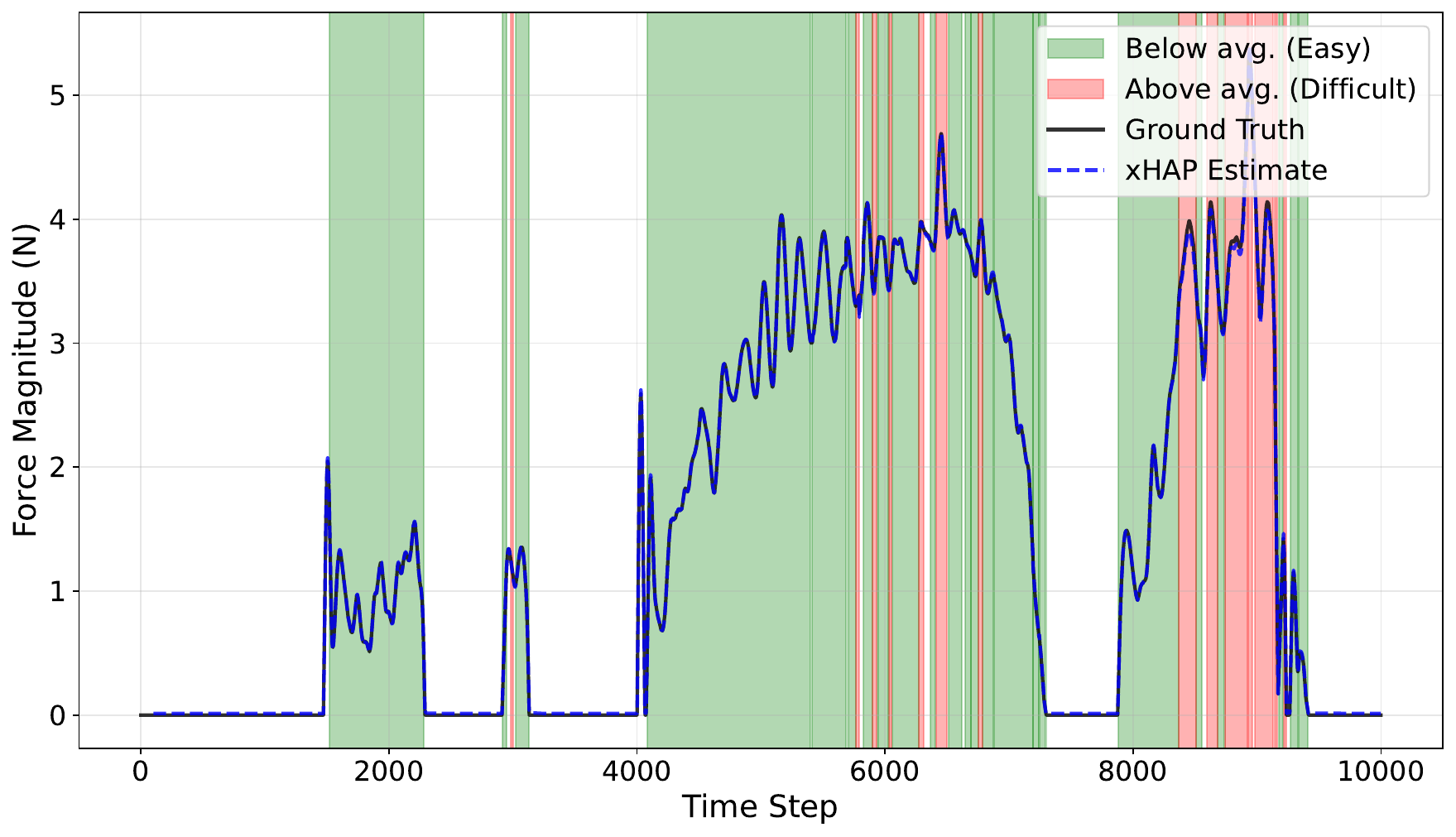}
 \caption{xHAP estimation of the haptic trace is divided in difficult and easy to estimate regions.}
 \label{fig:easy_hard}
 \vspace{-1em}
\end{figure}

\subsection{Haptic feature comparison}

\begin{figure}[!t]
 \centering
 \includegraphics[width=1\columnwidth]{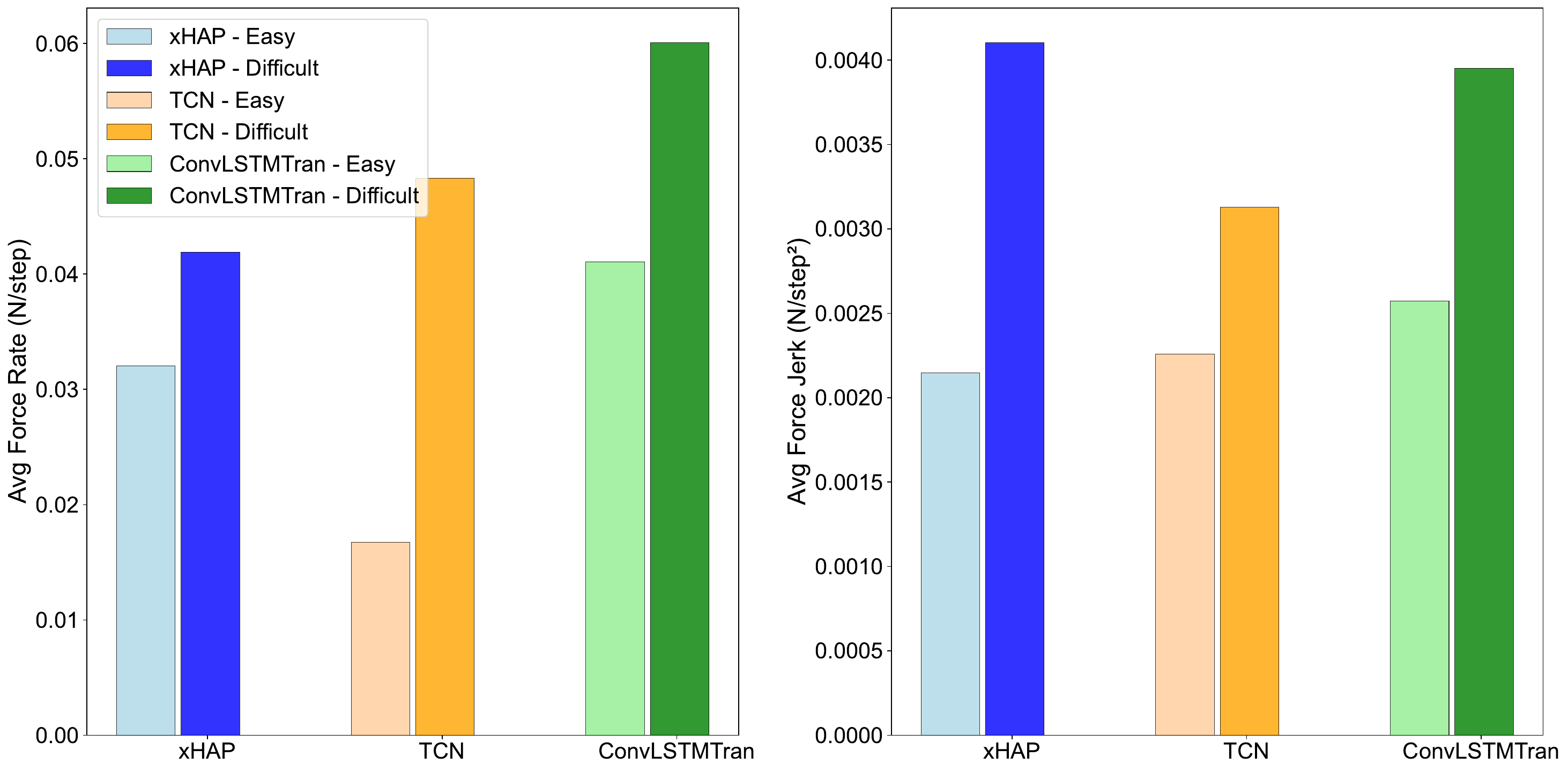}
 \caption{Comparison of average force rate and jerk across models, showing consistent separation of easy vs.\ difficult interactions.}
 \label{fig:force_dynamics}
 \vspace{-1em}
\end{figure}

Valuable insights can be gained by analyzing the different features of the haptic time series. Fig.~\ref{fig:easy_hard} shows that the estimation is divided into regions that are either easy or difficult to estimate. Fig.~\ref{fig:force_dynamics} demonstrates consistent results across the three evaluated models, with clear separation between easy and difficult regions in both average force rate and jerk, where rate and jerk refer to the first and second order time derivatives of force, respectively, which quantify the rate of change of force dynamics. For instance, in the xHAP and TCN models, the average force rate for difficult regions is approximately \(0.05\,\text{N/step}\), nearly double that of easy regions (\(\sim 0.025\,\text{N/step}\)). Similarly, the average force jerk in difficult regions remains below \(0.004\,\text{N/step}^2\) across all models, while easy regions cluster around lower values (\(<0.002\,\text{N/step}^2\)). This consistency indicates that dynamic features such as rate and jerk provide stable and discriminative cues for distinguishing material stiffness.

%---------------------- Results and Discussion -----------------------------------------------
\begin{table}[ht]
  \caption{Simulation and channel parameters}
  \label{tab:sim_channel_params}
  \centering
  \renewcommand{\arraystretch}{1.15}
  \resizebox{\columnwidth}{!}{%
  \begin{tabular}{ll}
    \hline
    \textbf{Description} & \textbf{Value} \\
    \hline
    \multicolumn{2}{l}{\textbf{Channel parameters}} \\
    \hline
    Carrier frequency & 1.8~GHz \\
    System bandwidth & 20~MHz \\
    Transmit power $P_{\mathrm{tx}}$ & 43~dBm \\
    Antenna gains ($G_{\mathrm{tx}}/G_{\mathrm{rx}}$) & 8~dBi / 0~dBi \\
    Receiver noise figure & 7~dB \\
    Receiver noise floor $N_{\mathrm{dBm}}$ & $-90.0$~dBm (20~MHz,\ 7~dB NF) \\
    Antenna heights (BS/UE) & 25~m / 1.5~m \\
    Path loss model & 3GPP UMa LOS/NLOS (TR~38.901--based~\cite{3gpp38901}) \\
    Shadowing (UMa) & $\sigma_{\mathrm{LOS}}{=}4$~dB,\ $\sigma_{\mathrm{NLOS}}{=}6$~dB \\
    Temporal correlation ($\rho$) & 0.95 \\
    Fading model & Rayleigh (default) \\
    Channel Diversity $L_{\mathrm{div}}$ & 3 \\
    Modulation and coding & QPSK,\ $R{=}0.602$ \\
    \hline
    \multicolumn{2}{l}{\textbf{Simulation parameters}} \\
    \hline
    Restoration threshold & 0.1~N (default) \\
    Target effective PLR & $10^{-5}$ \\
    Simulation steps & $10^6$ \\
    Packet size & 256~bits \\

    \hline
  \end{tabular}%
  }
  \vspace{-0.6em}
\end{table}

\section{\rev{Network Performance Evaluation}}
\label{sec:results}
This section presents a comprehensive evaluation of the proposed xHAP estimator in terms of communication reliability, coverage, and network capacity. We analyze the model’s performance under realistic channel conditions, comparing it against baseline, i.e. no haptic data restoration, and competing DL architectures. The results demonstrate how integrating haptic packet-loss restoration into the communication pipeline reduces SNR requirements, extends coverage, and enhances overall network capacity while maintaining stringent low-latency and reliability constraints.

\begin{figure}[!t]
 \centering
 \includegraphics[width=1\columnwidth]{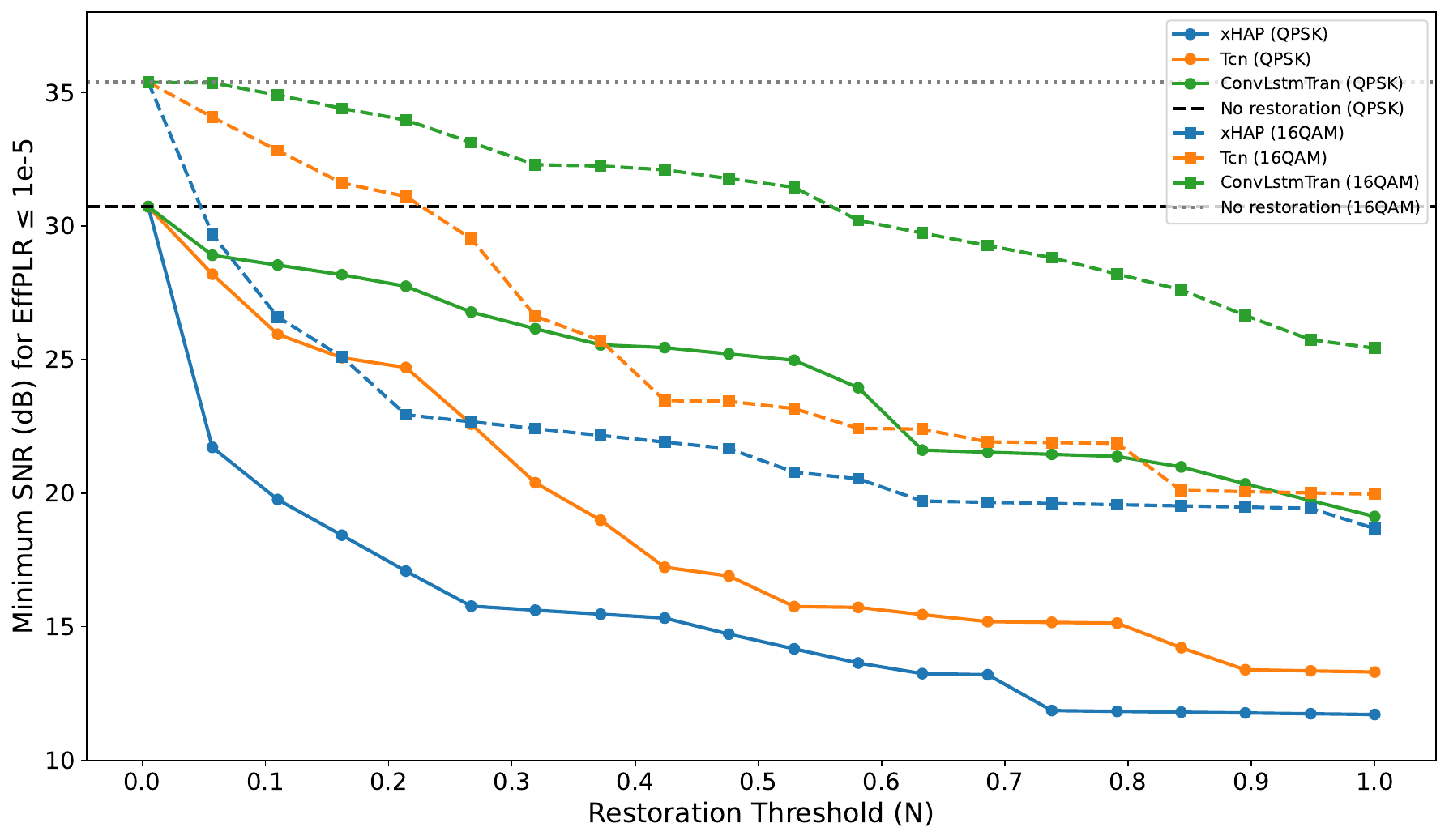}
     \caption{Minimum SNR for targeted reliability rate for QPSK, Coding rate $R=0.602$, and 16QAM, Coding rate $R=0.658$.}
 \vspace{-1em}
\label{fig:snr_vs_threshold}
\end{figure}

\begin{figure}[!t]
 \centering
 \includegraphics[width=1\columnwidth]{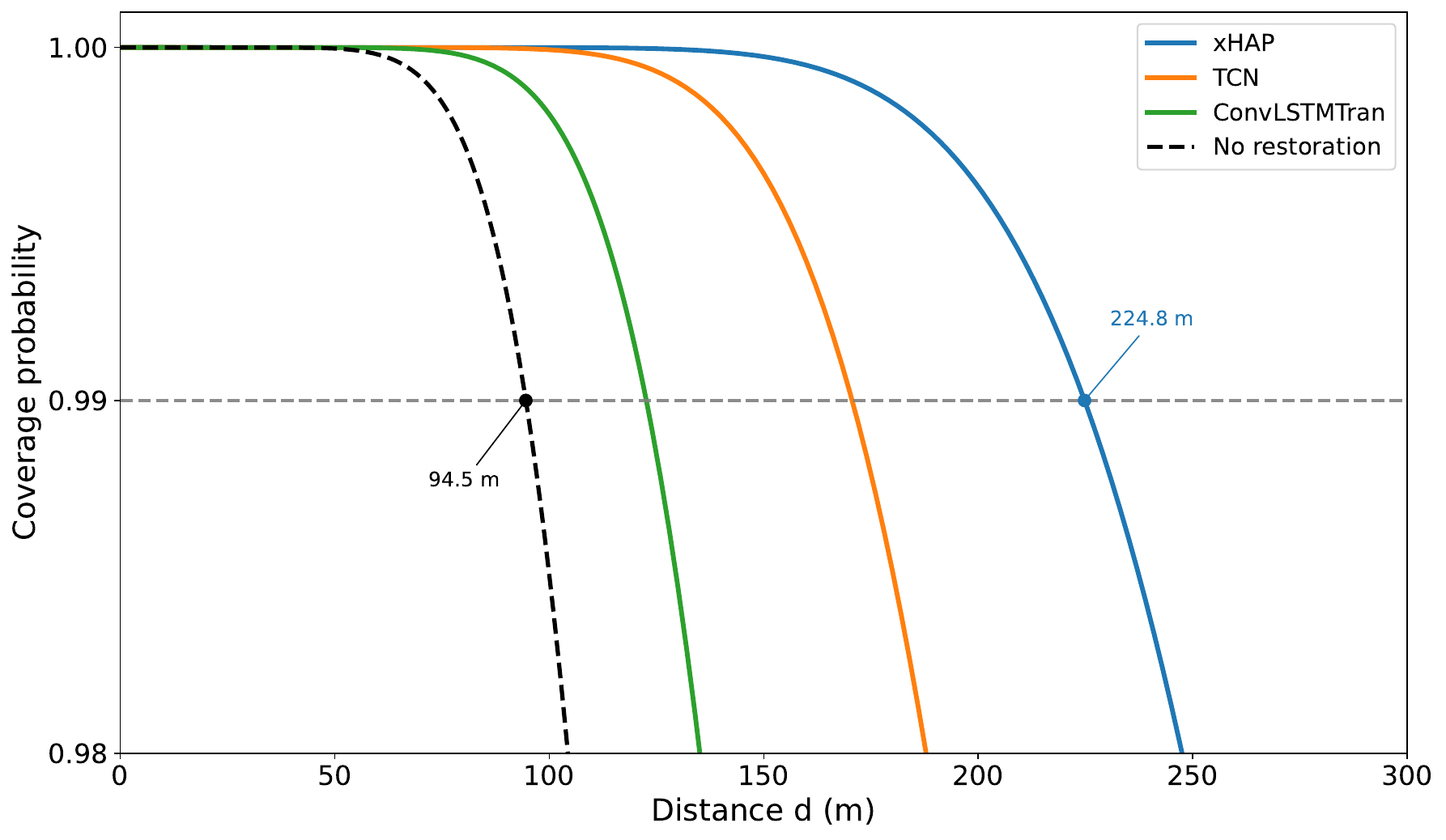}
     \caption{Coverage probability with restoration threshold $= 0.1$ for QPSK, Coding rate $R=0.602$.}
 \vspace{-1em}
\label{fig:cell-edge}
\end{figure}

\subsection{Coverage Distance and SNR}

Fig.~\ref{fig:snr_vs_threshold} illustrates the minimum SNR required to achieve a target effective PLR of $10^{-5}$ under various restoration thresholds $T_{\mathrm{thr}}$. The simulation step size is set to $10^6$, combining all test data trace activities shown in Table~\ref{tab:model_performance}. The performance evaluation, conducted using a binary search for the required SNR, compares the models against a baseline no-restoration scenario. At $T_{\mathrm{thr}}{=}0.1$, the baseline achieves the target reliability rate at minimum SNR of $30.82~\mathrm{dB}$ under QPSK, with  $R = 0.602$. Compared to this, restoration with xHAP achieves the target reliability with a minimum SNR of $19.91~\mathrm{dB}$, improving the SNR requirement by more than $10~\mathrm{dB}$. As we relax the error threshold, we can restore more packets with xHAP and operate at even lower SNR, hence providing higher SNR gain. In comparison, the other architectures yield more limited gains. The TCN model shows a moderate reduction in required SNR at $26.9~\mathrm{dB}$. The ConvLSTMTran model shows the least improvement, with its required SNR set at $28.74~\mathrm{dB}$.

\begin{figure*}[t]
  \centering
  \subfloat[]{\includegraphics[width=0.46\textwidth]{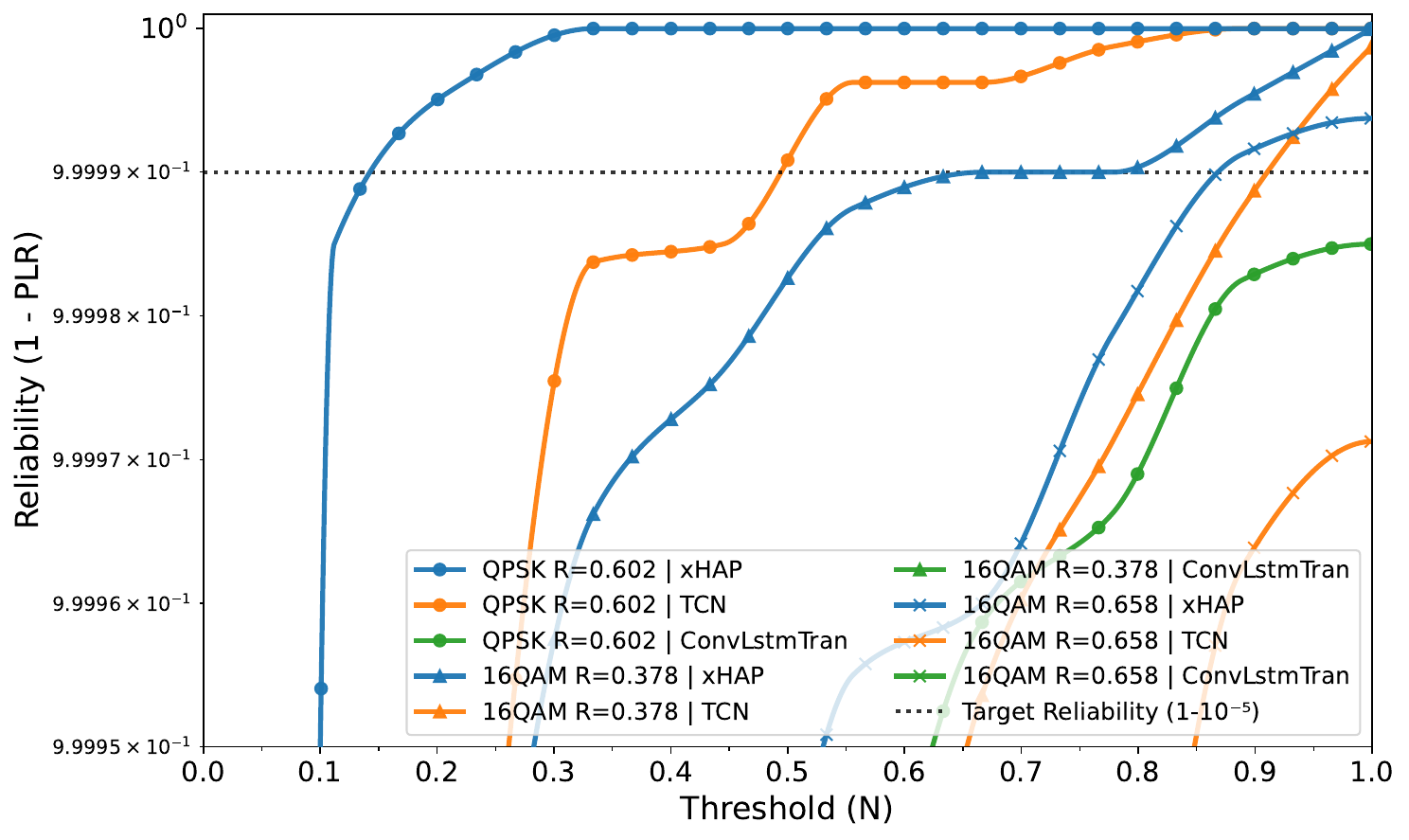}}\hfill
  \subfloat[]{\includegraphics[width=0.46\textwidth]{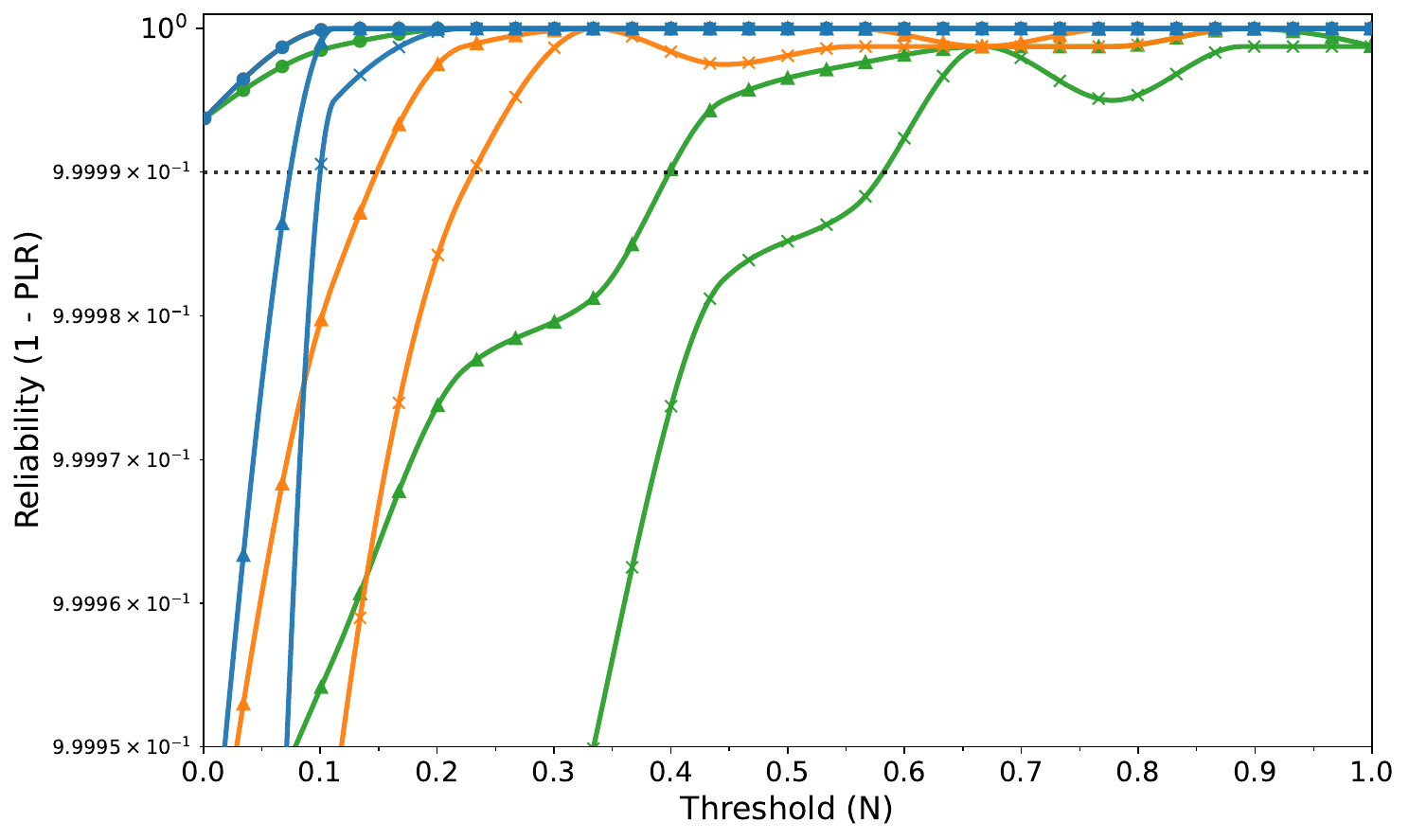}}\\[0.5em]
  \caption{\rev{Reliability $(1-\mathrm{PLR})$ versus restoration threshold for different modulation and coding schemes. 
  (a)~SNR~$=20$~dB; (b)~SNR~$=30$~dB.}}
  \label{fig:MCS_all}
\end{figure*}

\begin{figure*}[!t]
  \centering
  \subfloat[]{\includegraphics[width=0.46\textwidth]{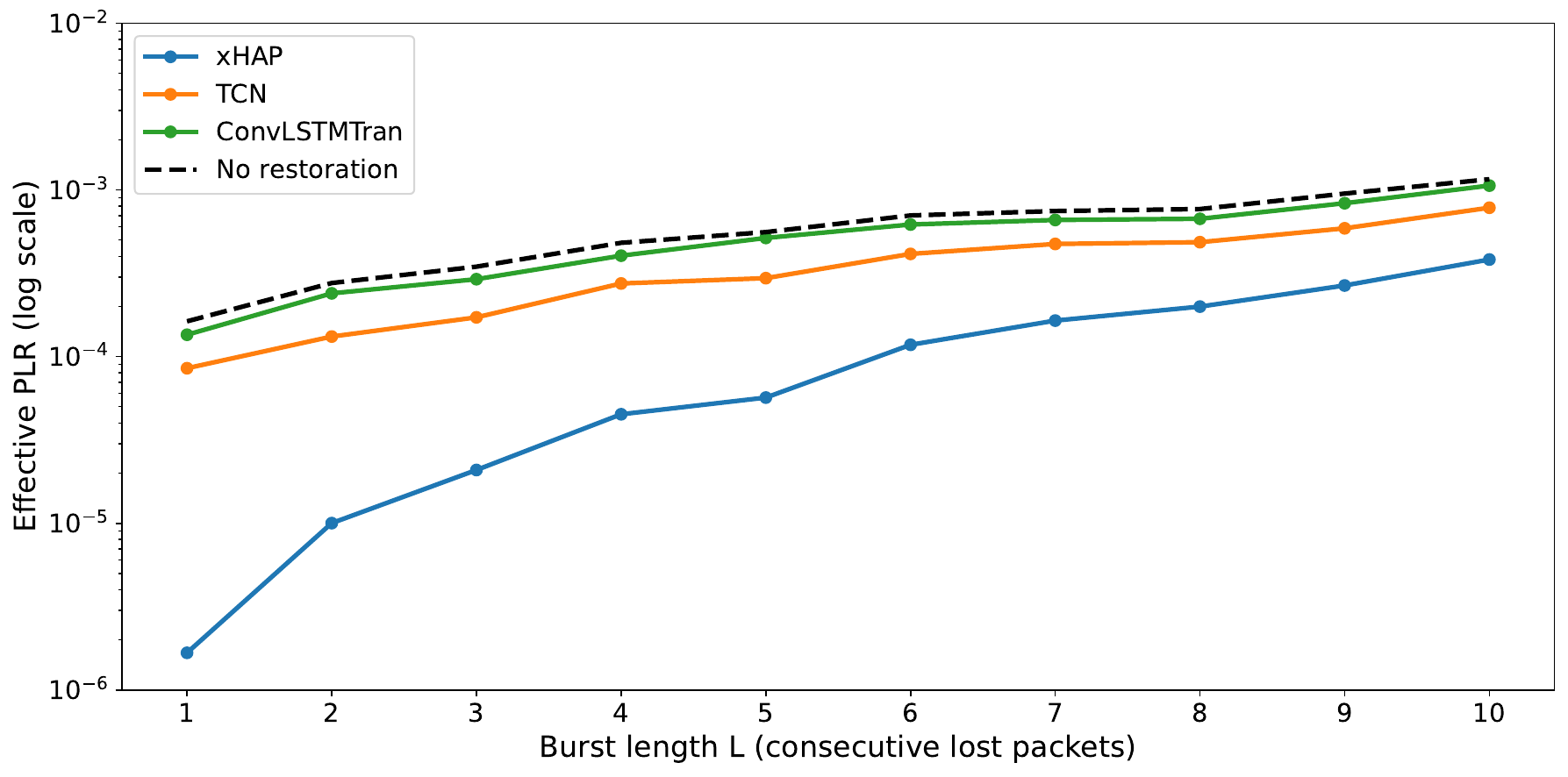}}\hfill
  \subfloat[]{\includegraphics[width=0.46\textwidth]{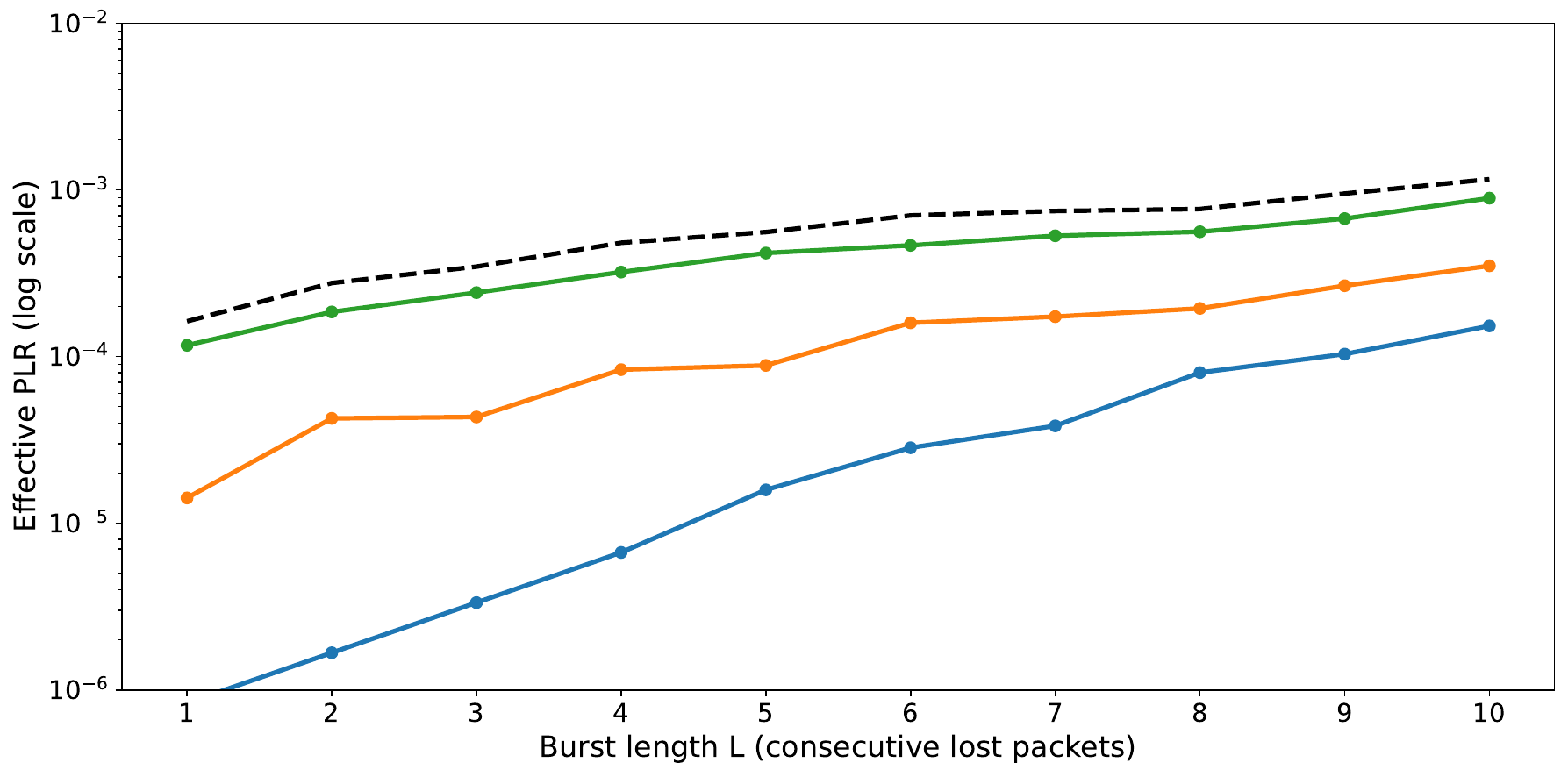}}\\[0.5em]
  \subfloat[]{\includegraphics[width=0.46\textwidth]{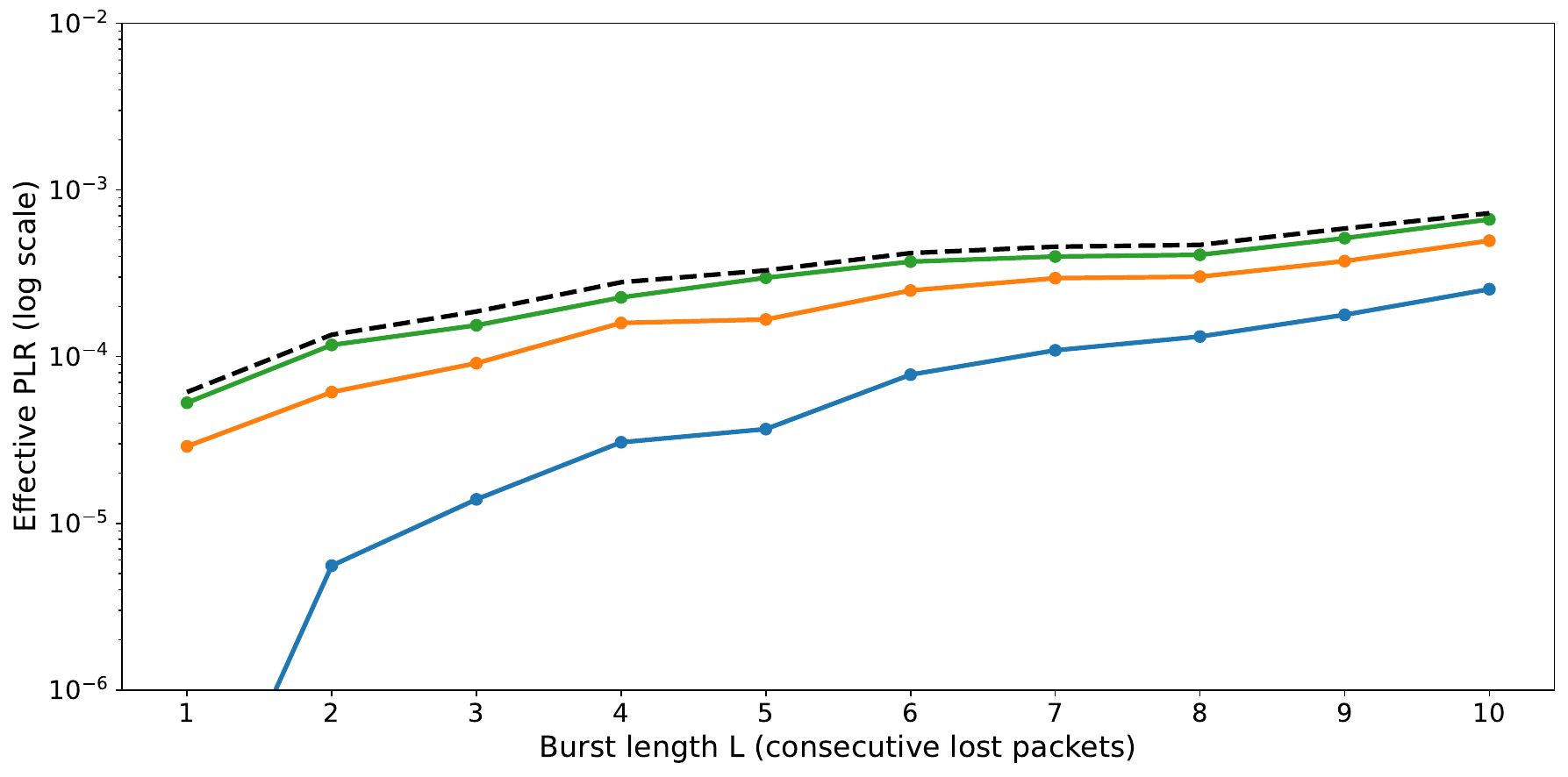}}\hfill
  \subfloat[]{\includegraphics[width=0.46\textwidth]{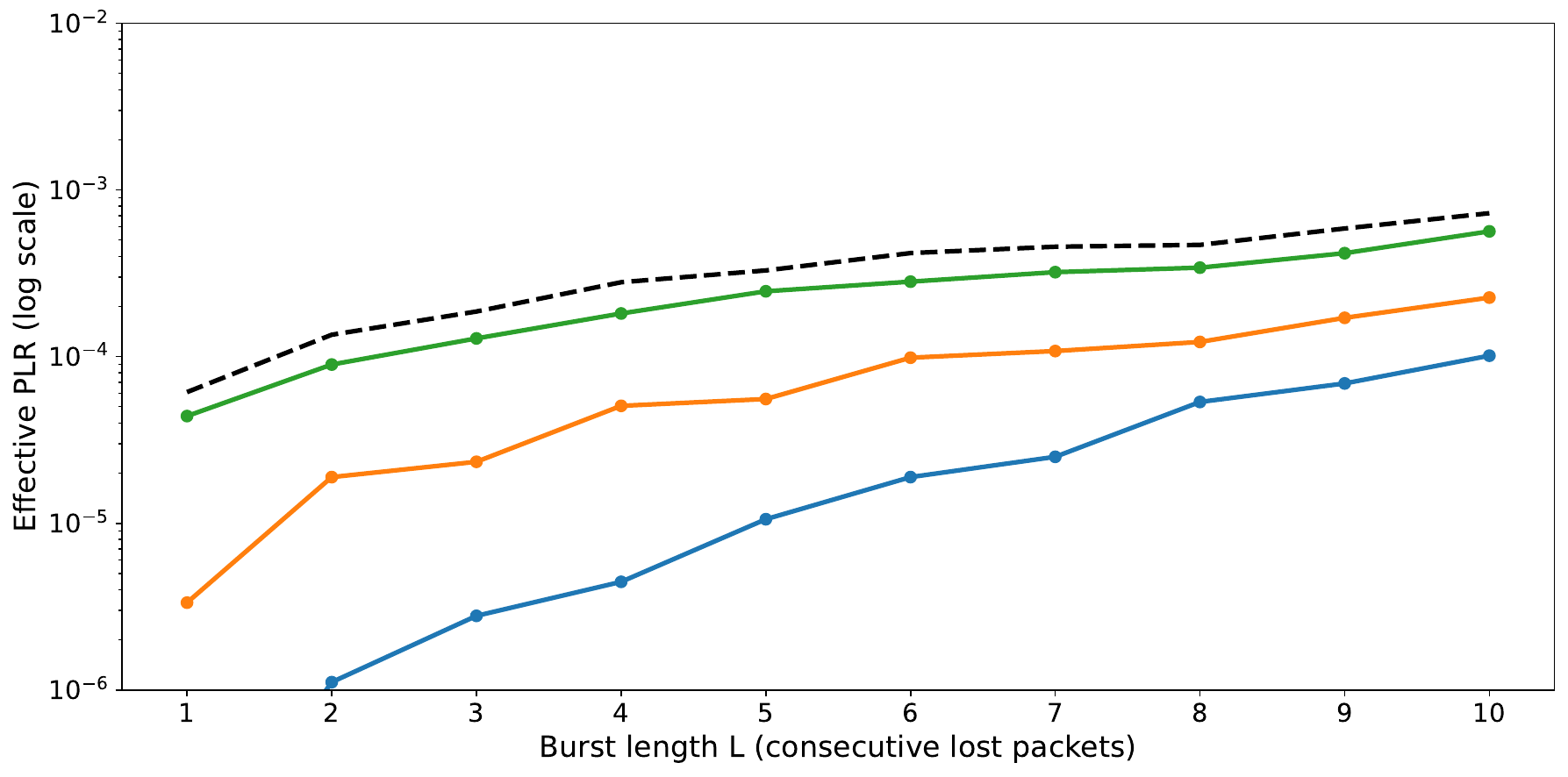}}

    \caption{Packet loss rate over increasing number of consecutive errors. 
    (a)~SNR~$=20$~dB, $T_{\mathrm{thr}}=0.1$; 
    (b)~SNR~$=20$~dB, $T_{\mathrm{thr}}=0.2$; 
    (c)~SNR~$=30$~dB, $T_{\mathrm{thr}}=0.1$; 
    (d)~SNR~$=30$~dB, $T_{\mathrm{thr}}=0.2$.}
    \label{fig:burst_error}
\end{figure*}

Another interesting result stems from the change of SNR for a higher Modulation and Coding Scheme (MCS). We set the MCS to 16QAM and $R = 0.658$, and observe that at $T_{\mathrm{thr}}{=}0.1$, xHAP requires $6~\mathrm{dB}$ higher SNR to achieve the same reliability, which also improves the data rate if required by the application.

For $T_{\mathrm{thr}}{=}0.1$, we evaluate the target coverage probability as a function of the cell-edge distance, as shown in Fig.~\ref{fig:cell-edge}. 
For coverage probability of $p_{\mathrm{cov}}{=}0.99$, xHAP-based restoration extends the coverage distance to $224.8~\mathrm{m}$, compared to $94.5~\mathrm{m}$ for the no-restoration baseline, thus increasing the coverage distance by 138\%, with ConvLSTMTran and TCN achieving intermediate ranges.

\subsection{Reliability vs. estimation error}

The restoration of lost packets is crucial for improving wireless link adaptation. This capability allows for the use of higher MCS indices, which in turn boosts spectral efficiency and data rates.

Figure~\ref{fig:MCS_all} compares the reliability performance for varying restoration error threshold across three different MCSs, for QPSK at $R = 0.602$ and 16QAM at $R = 0.378$ and $R = 0.658$, for SNR levels of $20~\mathrm{dB}$ and $30~\mathrm{dB}$. At $\mathrm{SNR}{=}20~\mathrm{dB}$, xHAP meets the reliability target at $T_{\mathrm{thr}}{\approx}0.1$, reducing the required threshold by roughly $0.4$ compared to TCN, while the ConvLSTMTran method is unable to meet the reliability target. At $\mathrm{SNR}{=}30~\mathrm{dB}$ with 16QAM and $R = 0.658$, xHAP can reach the reliability target near $T_{\mathrm{thr}}{=}0.1$, while TCN and ConvLSTMTran still require around $0.2$ and $0.6$, respectively. Ultimately, at the lower SNR level we require QPSK for reliable transmission with xHAP at $T_{\mathrm{thr}}{=}0.1$, but at $\mathrm{SNR}{=}30~\mathrm{dB}$, all three tested MCSs under xHAP meet the target reliability within $T_{\mathrm{thr}}{=}0.1$, highlighting the robustness of xHAP under favorable link conditions.

\begin{figure}[ht]
 \centering
 \includegraphics[width=1\columnwidth]{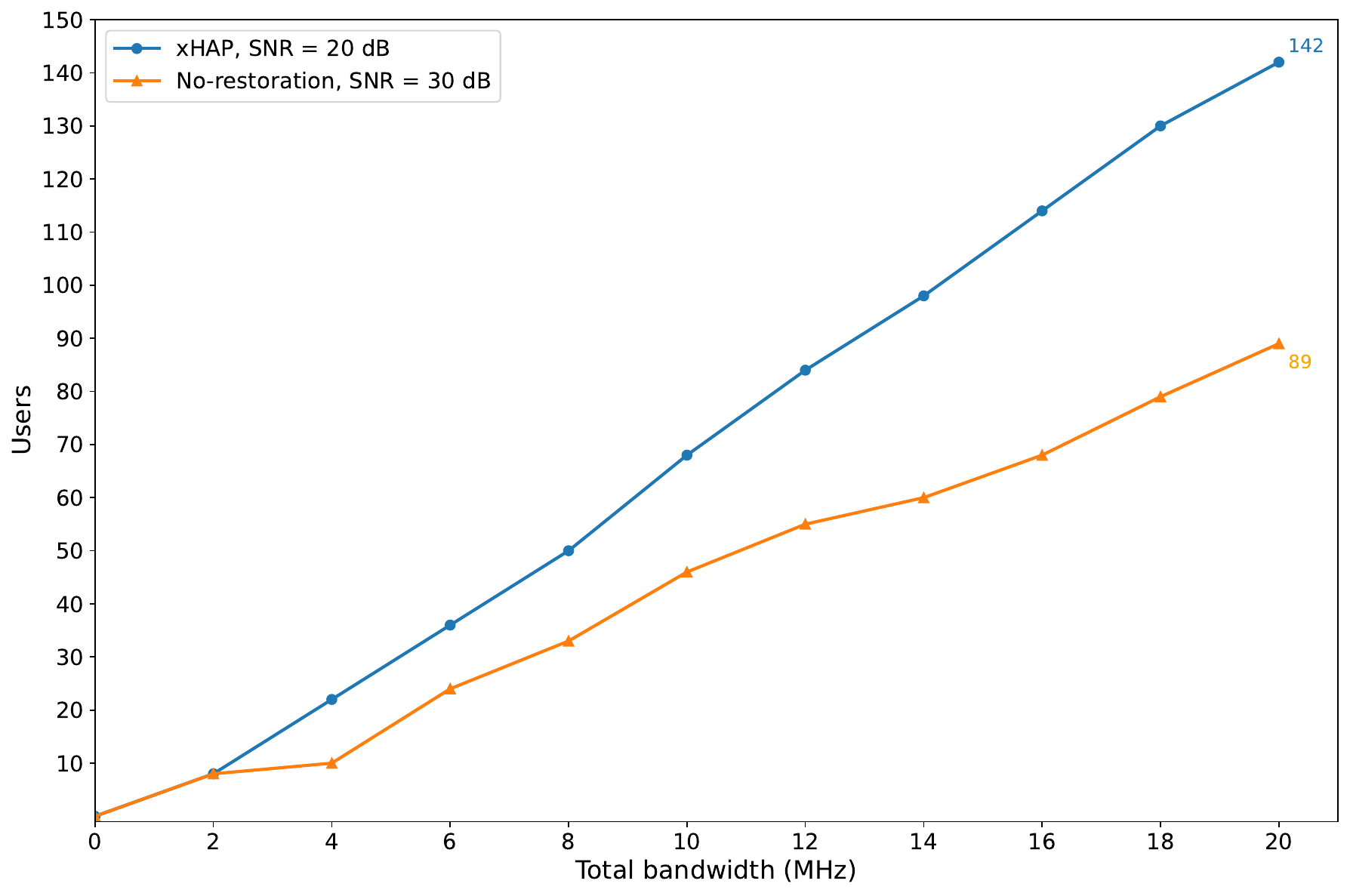}
 \caption{Number of admitted users over an increasing network bandwidth.}
 \vspace{-1em}
 \label{fig:network_capacity}
\end{figure}

\subsection{Consecutive burst error}
In scenarios where the channel coherence time spans multiple transmission time intervals (TTIs), burst errors may occur, resulting in consecutive packet losses. Figure~\ref{fig:burst_error} illustrates the variation in the effective packet loss rate (PLR) as a function of burst length under different signal-to-noise ratio (SNR) and threshold configurations. 

At an SNR of $20~\mathrm{dB}$, the results show that all models experience rapid degradation in reliability as the burst length increases. Under the stringent threshold of $T_{\mathrm{thr}}{=}0.1$, xHAP allows the system to operate only under single-packet losses, while none of the other models meet the target reliability at this SNR. When the threshold is relaxed to $T_{\mathrm{thr}}{=}0.2$, the proposed model demonstrates a substantial improvement, maintaining the target reliability for up to four consecutive lost packets, whereas the competing methods exhibit significantly higher effective PLR. Thus, at $20~\mathrm{dB}$, regardless of the threshold, the other methods fail to meet the required reliability.

At an SNR of $30~\mathrm{dB}$, we observe a better packet loss rate across all models. Under the default $T_{\mathrm{thr}}{=}0.1$, the xHAP model sustains the reliability requirement for up to two consecutive packet losses, outperforming the alternative approaches. When $T_{\mathrm{thr}}$ is relaxed to $0.2$, xHAP further extends its tolerance to five consecutive losses, reflecting its superior capability in handling temporally correlated fading and error propagation.

Overall, these results demonstrate that xHAP consistently achieves enhanced robustness against burst errors, maintaining reliability across a wider range of conditions compared to conventional temporal restoration models. Its performance scales more gracefully with increasing SNR and relaxed restoration error thresholds, confirming the effectiveness of the proposed design in mitigating deep-fade-induced packet losses.

\subsection{Network Capacity}

We define the \emph{network capacity} as the number of haptic users that can be admitted to and reliably served by the network while meeting the target reliability requirements. In all experiments, we fix $T_{\mathrm{thr}} = 0.1$ and gradually increase the network bandwidth up to $\mathcal{B} = 20~\mathrm{MHz}$. Figure~\ref{fig:network_capacity} illustrates the evolution of capacity as the network bandwidth $\mathcal{B}$ increases, comparing the proposed xHAP method with the baseline scenario under two initial SNR conditions. Each simulation consists of $10^6$ time steps, with all users initialized at a fixed SNR while shadowing and fading effects are still applied. The modulation scheme is set to QPSK with a coding rate of $R = 0.602$, consistent with the previous experiments.

\noindent
At $\mathrm{SNR}=30\,\mathrm{dB}$, the no-restoration baseline satisfies the reliability target for only $89$ users, and thus cannot ensure robust reliability for all users. By contrast, the xHAP-integrated system preserves strong reliability even at a lower $\mathrm{SNR}=20~\mathrm{dB}$, outperforming the $30~\mathrm{dB}$ baseline and increasing network capacity by $59.6\%$. In other words, our approach serves $59.6\%$ more users while relaxing the SNR requirement by $10~\mathrm{dB}$.

\section{Conclusion}
\label{sec:conclusion}

In this work, we introduced xHAP, a cross-attention based haptic restoration framework designed for force estimation under unreliable wireless links. By combining temporal attention with lightweight autoregressive modeling, xHAP reconstructs missing force feedback using both historical force data and live operator motion cues. The proposed method achieves high restoration performance across a wide range of haptic activities, outperforming convolutional, recurrent, and hybrid baselines in both accuracy and computational efficiency. Despite its compact architecture, xHAP generalizes well to both dynamic and rigid-body interactions, showing that cross-attention mechanisms can capture transient dynamics more effectively than deeper or more complex models. Beyond model performance, we also thoroughly evaluate the contribution of haptic restoration to wireless communication. When included in the wireless control loop, xHAP reduces SNR requirements by $10.58~\mathrm{dB}$ compared to the baseline while maintaining sub-millisecond inference latency. These improvements directly enhance coverage, increasing the distance by $138\%$, and network capacity with up to $59.6\%$ higher user support under realistic 3GPP channel conditions operating at $10~\mathrm{dB}$ lower than the no-restoration scenario.

\rev{A remaining limitation is that the current traces were collected under similar acquisition conditions and no live hardware-in-the-loop user study is included. Future work will evaluate xHAP across different operators, devices, tasks, contact dynamics, and physical teleoperation setups.} Overall, this work shows that intelligent restoration at the application layer can improve both reliability and latency in future haptic communication systems. By combining lightweight cross-modal attention with channel-aware design, xHAP offers a scalable and perceptually stable solution for ultra-reliable haptic interaction in 5G and beyond.

\bibliographystyle{IEEEtran}
\bibliography{references}

\begin{IEEEbiography}[{\includegraphics[width=1in,height=1.25in,%
clip,keepaspectratio]{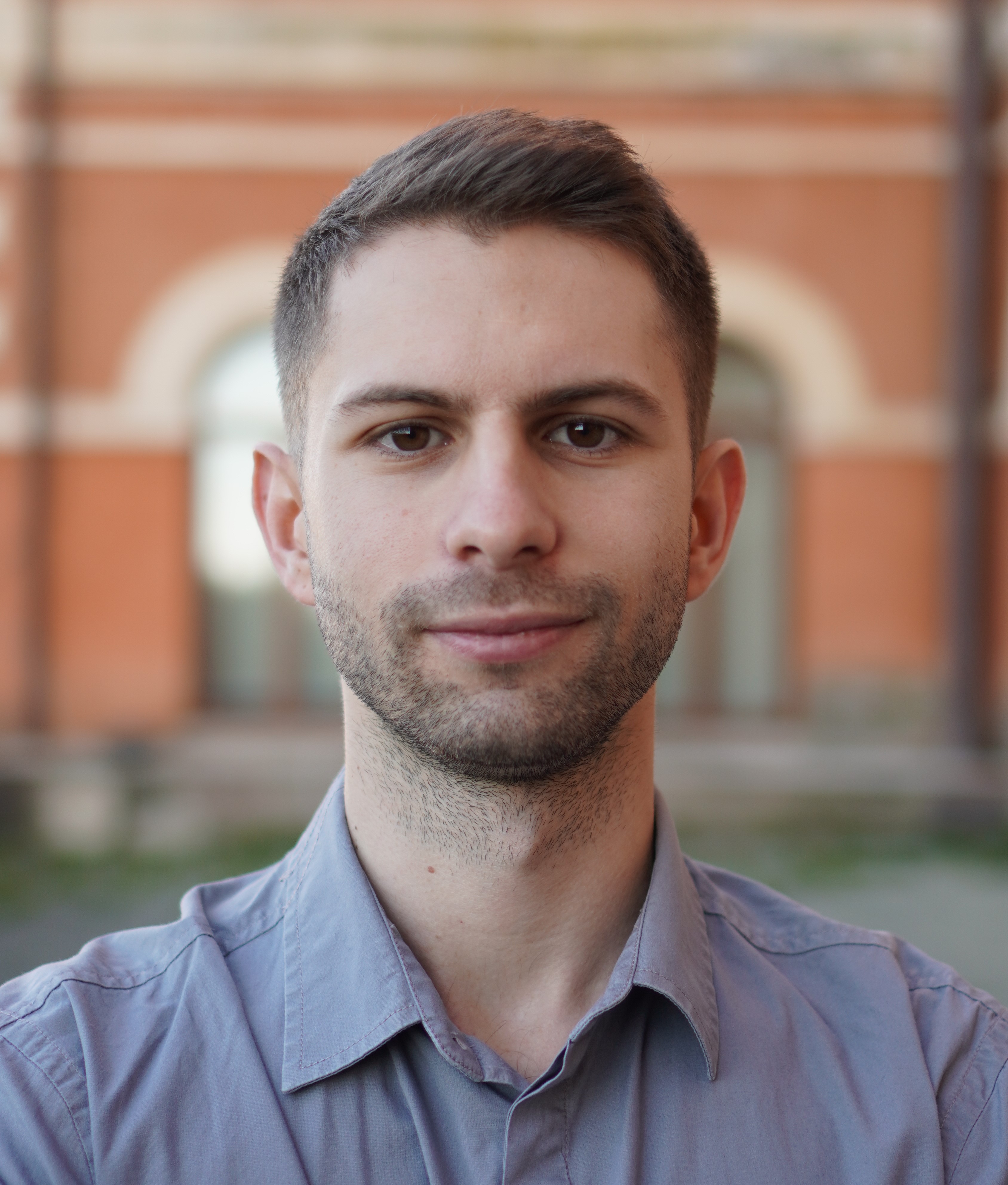}}]{Georgios Kokkinis (GS '25)}
 is a Ph.D. candidate in the Department of Electrical and Computer Engineering at Aarhus University, Denmark, as a Doctoral Candidate of the TOAST MSCA Doctoral Network. Previously, he was a researcher with the Department of Electrical and Computer Engineering, University of Huddersfield, U.K. He received the Diploma (five-year, integrated M.Eng.) in Electrical and Computer Engineering from Aristotle University of Thessaloniki, Greece. His research interests include machine learning for wireless communications, the Tactile Internet and haptic communications, wireless networks, signal processing, and programming.
\end{IEEEbiography}

\begin{IEEEbiography}[{\includegraphics[width=1in,height=1.25in,%
clip,keepaspectratio]{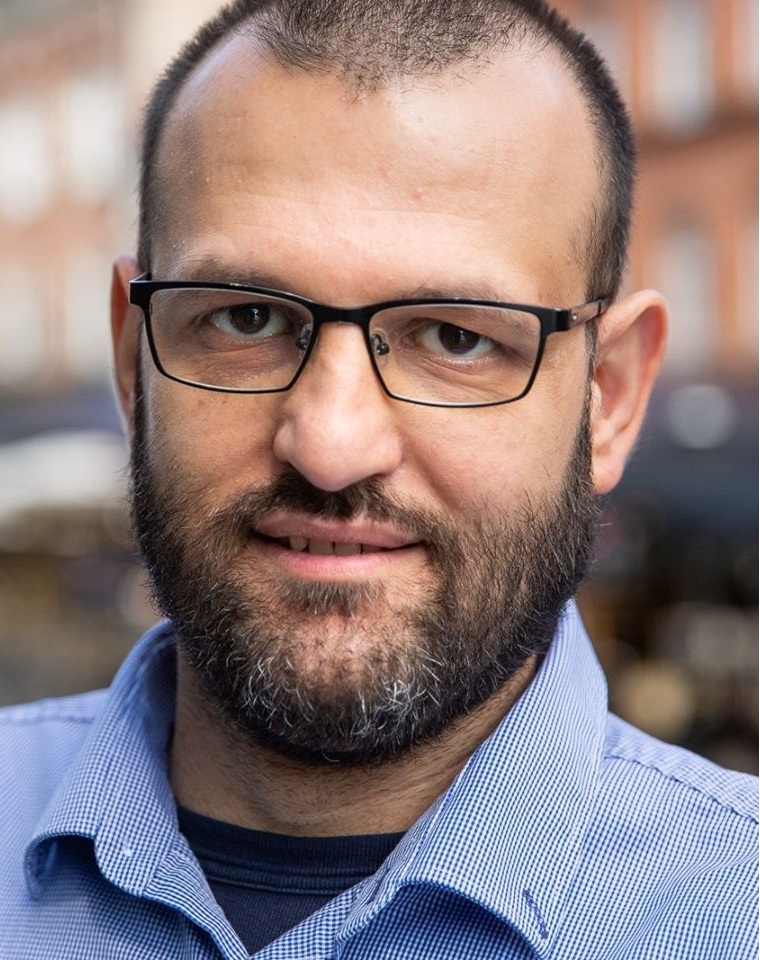}}]{Alexandros Iosifidis (SM'16)}
 is a Professor of Machine Learning at Tampere University, Finland, where he leads the Computational Intelligence group at the Unit of Computing Sciences, and the Fundamental Machine Learning research theme at the Data Science Research Centre. His research interests focus on topics of neural networks and statistical machine learning finding applications in computer vision, financial data and graph analysis problems. He is a member of the IEEE Technical Committee on Machine Learning for Signal Processing and the IEEE Technical Committee on Image, Video, and Multimedia Signal Processing. He served as the Associate Editor-in-Chief of Neurocomputing journal covering the neural networks research area between 2021 and 2025. He has served as Associate Editor of IEEE Transactions on Neural Networks and Learning Systems, IEEE Transactions on Artificial Intelligence, and IEEE Transactions on Circuits and Systems for Video Technology. He served as an Area Chair for IEEE ICIP 2018-2026 and IEEE ICASSP 2023-2025, as a Senior Area Chair of IEEE ICASSP 2026, and is a Virtual \& Web Chair of IEEE ICASSP 2029.
\end{IEEEbiography}

\begin{IEEEbiography}[{\includegraphics[width=1in,height=1.25in,%
clip,keepaspectratio]{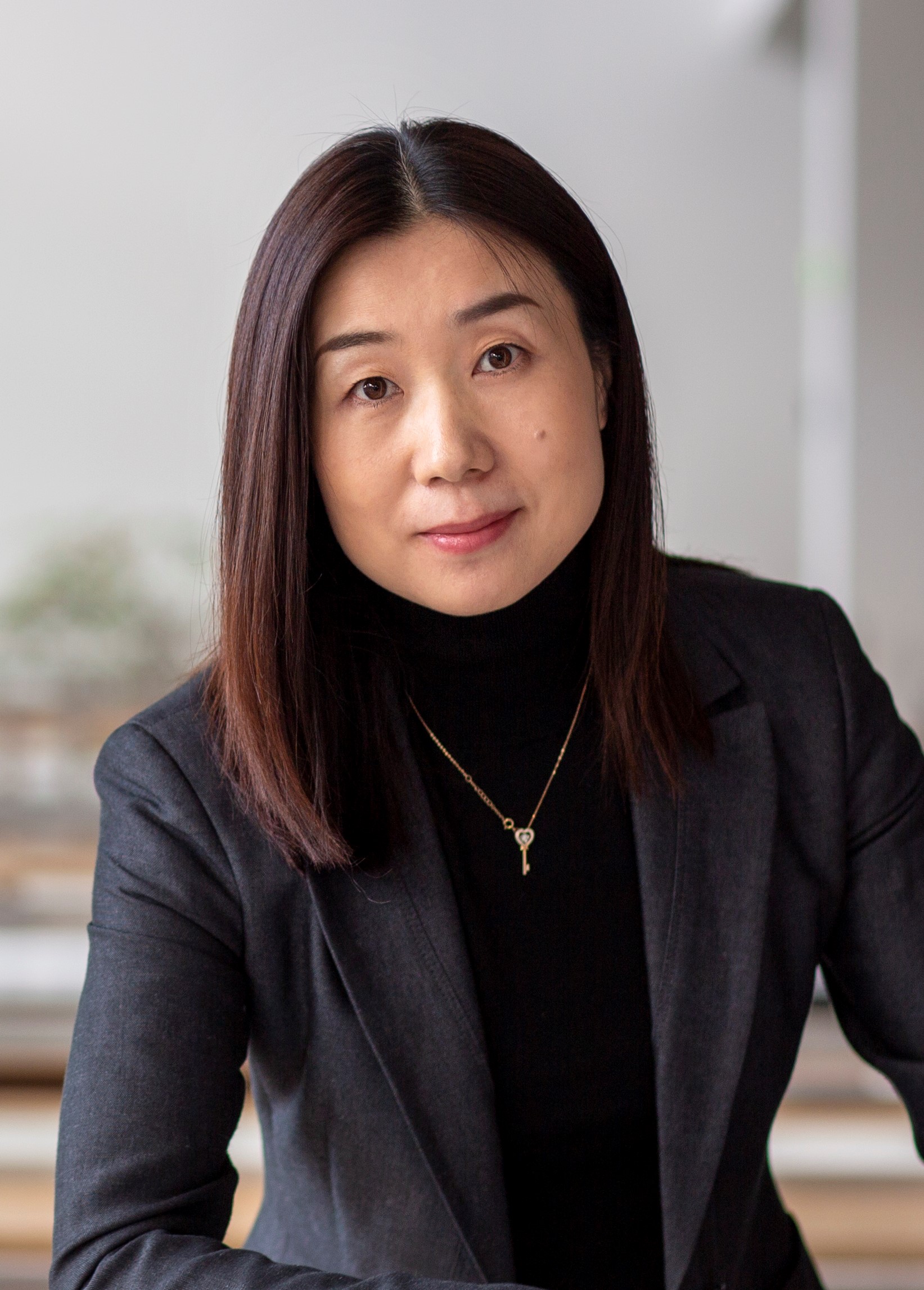}}]{Qi Zhang (SM '21)}
 is a Professor with the Department of Electrical and Computer Engineering, Aarhus University, Aarhus, Denmark. She is leading the Internet of Things research area of AU Research Centre for Digitalisation, Big Data and Data Analytics (DIGIT). Her research interests include Internet of Things, Edge Intelligence, Tactile Internet, Goal-oriented Semantic Communication, as well as sensor data compression and analytics. She is the PI of three Danish Independent Research Fund Projects: AgilE-IoT (Agile Edge Intelligence for Delay Sensitive IoT) and Light-IoT (Analytics Straight on Compressed IoT Data) and eTouch (Edge Intelligence for Immersive Telerobotics in Touch-enabled Tactile Internet). She is the project coordinator and PI of Horizon Europe MSCA Doctoral Networks TOAST (Touch-enabled Tactile Internet Training Network and Open Source Testbed). She was an Associate Editor of EURASIP Journal on Wireless Communications and Networking. She has (co-)authored more than 140+ publications in high impact journals and flagship conferences.
\end{IEEEbiography}

\end{document}